\documentclass{article}
\usepackage{graphicx}
\usepackage{amsfonts}
\usepackage{amsmath}
\usepackage{amssymb}
\usepackage{url}
\usepackage{bbm}
\usepackage{dsfont}
\usepackage{fancyhdr}
\usepackage{indentfirst}
\usepackage{enumitem}
\usepackage[normalem]{ulem}
\usepackage{mathrsfs}
\usepackage{multirow} 
\usepackage{diagbox}
\usepackage{appendix}
 \usepackage[colorlinks=true,citecolor=blue]{hyperref}
\usepackage{amsthm}
\usepackage{color}

\definecolor{DarkGreen}{rgb}{0.2,0.6,0.2}

\def\blue#1{\textcolor{blue}{#1}}

\definecolor{purple}{rgb}{0.6,0.3,0.8}

\usepackage{natbib}
\usepackage{comment}

\def\d{\mathrm{d}}
\def\laweq{\buildrel \mathrm{d} \over =}

\newcommand{\var}{\mathrm{Var}}

\newcommand{\E}{\mathbb{E}}

\newcommand{\R}{\mathbb{R}}

\newcommand{\mG}{\mathcal{G}}

\newcommand{\RR}{\overline{\mathbb R}}

\newcommand{\p}{\mathbb{P}}
\newcommand{\X}{\mathcal{X}}

\newcommand{\G}{\mathcal{G}}

\newcommand{\Q}{\mathbb{Q}}

\newcommand{\id}{\mathds{1}}

\renewcommand{\ge}{\geqslant}
\renewcommand{\le}{\leqslant}
\renewcommand{\geq}{\geqslant}
\renewcommand{\leq}{\leqslant}
\renewcommand{\epsilon}{\varepsilon}

\newcommand{\essinf}{\mathrm{ess\mbox{-}inf}}
\renewcommand{\cdots}{\dots}
\theoremstyle{plain}
\newtheorem{theorem}{Theorem}

\newtheorem{lemma}{Lemma}
\newtheorem{proposition}{Proposition}
\theoremstyle{definition}
\newtheorem{definition}{Definition}
\newtheorem{example}{Example}

\theoremstyle{remark}
\newtheorem{remark}{Remark}

\theoremstyle{plain}

\theoremstyle{definition}

\theoremstyle{remark}

\usepackage[misc]{ifsym}

\usepackage[onehalfspacing]{setspace}

\usepackage{centernot}

\newcommand{\ES}{\mathrm{ES}}

\usepackage{footmisc}
\begin{document}

\title{Jensen gaps and the decomposition principle}
\author{Lilian Hu\thanks{Department of Statistics and Actuarial Science, University of Waterloo,  Canada. \Letter~{\scriptsize\url{l66hu@uwaterloo.ca}}}
\and Ruodu Wang\thanks{Department of Statistics and Actuarial Science, University of Waterloo,  Canada. \Letter~{\scriptsize\url{wang@uwaterloo.ca}}}}
\date{\today}
 
\maketitle

\begin{abstract}
% We provide axiomatic characterizations of the Jensen-gap functionals of the form
% $
%     \rho(X)=\E[f(X)]-f(\E[X]),
% $
% where $f$ is a real-valued function called the generator of the Jensen gap. We first formulate conditional versions of law-invariant functionals. 
% % On an atomless probability space, we show that, the decomposition principle that generalizes the law of total variance, together with the law-invariance and some basic conditions, forces the Jensen-gap representation. We establish the result through two approaches, starting respectively from general simple random variables and from equiprobable random variables with $2^n$ equally likely states. Under suitable continuity assumptions, the representations extend to broader classes of random variables, including integrable random variables. 
% On an atomless probability space, we obtain two characterizations
% of Jensen-gap functionals: one on arbitrary simple random variables and the other
% on dyadic equiprobable random variables. In both settings, law invariance and the
% corresponding decomposition principle are sufficient. Under suitable additional
% continuity and regularity assumptions, both characterizations extend to broader
% classes, including integrable random variables. Finally, additional homogeneity
% or invariance properties identify logarithmic, $x\log x$, and power generators.
% % Our results separate the role of conditional decomposition, which determines the general Jensen-gap structure, from that of additional invariance properties, which determine the specific generator.
We provide axiomatic characterizations of Jensen-gap functionals, which have the form
\(
\rho(X)=\E[f(X)]-f(\E[X])
\)
for some real-valued function $f$ called the generator. This class includes important examples such as the excess growth rate, the utility premium, the Kullback–Leibler (KL) divergence, and the variance, and is also closely related to measures of variability. 
We analyze various forms of convexity for Jensen gaps. 
To characterize Jensen gaps, we formulate conditional versions of law-invariant functionals,
which are needed to state the decomposition principle. On an atomless
probability space, we show that law invariance together with a binary
decomposition principle is sufficient to characterize Jensen gaps on simple random variables. Under
suitable continuity conditions, the characterization extends to broader
classes, including compactly supported random variables and integrable random variables. We then identify
the generator $f$ by imposing either of two natural invariance conditions:
scaling invariance or translation invariance. The former identifies logarithmic, \(x\log x\), and power generators, whereas the latter identifies exponential and quadratic generators.

\medskip \noindent \textbf{Keywords}: Jensen gap, decomposition principle,  conditional risk measures, measures of variability, excess growth rate
\end{abstract}

\section{Introduction}

For a real-valued function $f$  on a subset of $\R$, define the functional $J_f$  by 
\begin{equation}\label{eq:jensen_gap_form}
    J_f(X)
    =
    \E[f(X)]-f(\E[X]),
\end{equation}
on a suitable domain of random variables $X$ such that the
  right-hand side of \eqref{eq:jensen_gap_form} is well-defined. 
We call the functional $J_f$ in \eqref{eq:jensen_gap_form} a 
\emph{Jensen gap} generated by $f$, which is motivated by Jensen's inequality:  $J_f(X)\geq0$ whenever $f$ is convex.\footnote{
This terminology is used, for instance, by \citet{GSR19}.}
It quantifies the effect of the dispersion of $X$ through the curvature of $f$. This effect depends jointly on
the distribution of $X$ and the curvature of $f$; affine components of $f$
do not contribute to the gap.
We do not assume $f$ is convex in the definition, although the convex case is the most interesting.

Jensen gaps arise naturally in finance and economics. An important example is the excess
growth rate \citep{campbell2025excessgrowth}. Consider a portfolio with
weights $\pi=(\pi_1,\ldots,\pi_n)$  in the standard simplex and positive gross asset returns
$R=(R_1,\ldots,R_n)$. Its excess growth rate is
 $$
    \Gamma(\pi,R)
    =
    \log\left(\sum_{i=1}^n\pi_iR_i\right)
    -
    \sum_{i=1}^n\pi_i\log R_i.
 $$
% If $\widetilde R$ satisfies
% $\mathbb P(\widetilde R=R_i)=\pi_i$, then
% \[
%     \Gamma(\pi,R)
%     =
%     \E[-\log\widetilde R]
%     +
%     \log\E[\widetilde R]
%     =
%     J_{-\log}(\widetilde R).
% \]
Hence, the excess growth rate is related to the Jensen gap
generated by $f=-\log$. Other examples include several common measures of variability. When
$f(x)=x^2$, $J_f(X)$ is the variance of $X$. For $f(x)=x^3$ or $x^4$,
the corresponding Jensen gaps are related to higher-order features of the
distribution, such as skewness and kurtosis. Moreover, if
$f(x)=(\max(x-\tau,0))^2$, then
\(
J_f(X)
=
\E[(\max(X-\tau,0))^2]
-
(\max(\E[X]-\tau,0))^2.
\)
When $\E[X]\leq\tau$, this reduces to the threshold upper semi-variance of
$X$. More generally, if a nonnegative Jensen gap additionally
satisfies homogeneity of some degree $\alpha\geq0$, then it belongs to the class
of variability measures considered by \citet{bellini2022parametric}.
% Another fundamental example is the variance:
% when $f(x)=x^2$,
% \[
%     J_f(X)
%     =
%     \E[X^2]-\E[X]^2
%     =
%     \mathrm{Var}(X).
% \]
% This also connects Jensen gaps with variability measures; see, for example,
% \citet{bellini2022parametric}.
  Jensen gaps also arise in derivative pricing. Suppose
that a  positive forward price satisfies
\(
\d F_t=F_t\sigma_t\,\d W_t
\)
under the risk-neutral measure $\mathbb Q$. Under suitable regularity and
integrability conditions, It\^o's formula gives
 $$
\E^{\mathbb Q}[f(F_t)]-f(\E^{\mathbb Q}[F_t])
=
\E^{\mathbb Q}[f(F_t)]-f(F_0)
=
\frac 12 \E^{\mathbb Q}
\left[\int_0^t F_s^2f''(F_s)\sigma_s^2\,\d s\right].
 $$
Thus, the Jensen gap can be interpreted in terms of expected weighted
realized variance. In particular, for $f =-\log  $, the weighting term $F_s^2f''(F_s)$ equals $1$, and the resulting quantity is closely related to the classical replication and pricing of variance swaps \citep{carrlee2009volatility}.

In addition, for a positive random variable \(X\) representing income, the mean log deviation \(J_{f}(X)\) with $f=-\log$ and the Theil index \(J_{f}(X)/\mathbb E[X]\) with $f(x)=x\log x$ are two classical measures of income inequality \citep{Shorrocks1980}. In expected utility theory, for a concave utility function \(u\) and random wealth $X$, the utility premium \(J_{-u}(X)=u(\mathbb E[X])-\mathbb E[u(X)]\) measures the loss in expected utility caused by uncertainty \citep{EeckhoudtSchlesinger2009}.

Beyond finance and economics, Jensen gaps with convex generators \(f\) are also related to important quantities in information theory and related areas.  For $f$ that is further differentiable, the quantity
\(
    B_f(x,y)
    =
    f(x)-f(y)-f'(y)(x-y)
\)
is called the Bregman divergence generated by $f$. Then
\(
    J_f(X)
    =
    \E[
        B_f (X,\E[X])
    ],
\)
so the Jensen gap can be viewed as the expected Bregman divergence
between a random outcome and its mean, a quantity often referred to as Bregman information \citep{bregman1967relaxation,banerjee2005clustering,gruber2023uncertainty}.
In addition, the Jensen gaps are also related to the $f$-divergence \citep{csiszar1967information,csiszar2008axiomatic}. The $f$-divergence of a probability distribution $P$ from $Q$ is $D_f(P||Q)=\E^Q[f(Z)]$, where $Z=\d P/\d Q$, assuming $P\ll  Q$. Hence, if the expectation is taken under the measure $Q$ and $f(1)=0$, then $J_f(Z)=\E^Q[f(Z)]=D_f(P||Q)$. Some common examples are the Kullback--Leibler (KL) divergence, obtained from $f(x)=x\log x$ (with $0\log0=0$), and the chi-square divergence, obtained from
$f(x)=(x-1)^2$.

A key feature of the  Jensen-gap functionals is the
\emph{decomposition principle}. For a sub-$\sigma$-algebra $\mathcal G$,
a Jensen-gap functional satisfies
 \begin{equation}
\rho(X)
=
\rho\bigl(\E[X\mid\mG]\bigr)
+
\E[\rho(X\mid\mG)],\label{eq:DP}
 \end{equation}
 whenever all terms are
well-defined. Here $\rho(X\mid\mG)$ is defined in a natural way similar to the conditional expectation. 
 In the case of a finite $\mG$, 
$ \rho(X\mid\mG)$
can be interpreted as the functional $\rho$ applied to the conditional distributions of $X$ given atoms in $\mG$. 
For a general $\mG$, a more careful analysis is needed to form a proper definition, which we study in  Section~\ref{s:cond_fun_general_G}. For now, the reader may keep the discrete case in mind for intuition. 
 The  construction of $ \rho(X\mid\mG)$ is closely related to conditional risk measures, studied by \cite{FrittelliRosazzaGianin2004}, \cite{DetlefsenScandolo2005}, and \cite{RuszczynskiShapiro2006}. 

To interpret 
the decomposition principle \eqref{eq:DP} in financial and economic applications, 
imagine that $\rho$ represents a measurement of variability and $X$ a random  potential loss.
The $\sigma$-algebra $\mathcal G$ models an economic environment. For instance, we can take $\mathcal G=\sigma(\theta)$ for a latent random factor $\theta$ such as a stochastic interest rate or a random economic state; see e.g., \cite{Gordy2003} in the context of credit risk.  
% In this model, $\E[X\mid \mathcal G]$ represents how the expected loss varies across different values of $\theta$, and 
% $\rho(X\mid \mathcal G)$ measures the the uncertainty within the scenario of each value of $\theta$.  
%Credit risk models distinguish these systematic and idiosyncratic sources of risk \citep{Gordy2003}. This distinction naturally motivates decomposing the overall variability of credit losses into two components. 
%Let \(X\) denote a credit loss and \(\mG\) represent information about the economic state.
The term \(\rho(\mathbb E[X\mid\mG])\) captures the variability of conditional expected losses across economic conditions, while \(\mathbb E[\rho(X\mid\mG)]\) captures the average variability remaining within each economic state. The decomposition principle \eqref{eq:DP} thus makes the explicit rule that these two sources of variability add up to the total variability. 
While such a rule is certainly restrictive, it is arguably the simplest way of summing up the sources, and it
is satisfied by many useful mappings, such as the variance, the KL divergence, and  the other quantities discussed above.  

Another example arises in the measurement of income inequality. If \(X\) represents individual income and \(\mG\) identifies geographic regions, income inequality can be decomposed into differences in average income between regions \(\rho(\mathbb E[X\mid\mG])\) and expected income inequality within each region \(\mathbb E[\rho(X\mid\mG)]\). Such between-group and within-group decompositions are widely studied in the measurement of income inequality \citep{Shorrocks1980}.
% The decomposition principle requires these two components to add up to the overall assessment.

More generally, these examples illustrate the importance of consistency in financial and economic evaluations across different information levels. A random outcome can be assessed directly from its unconditional distribution, or by first distinguishing different scenarios and then evaluating the remaining uncertainty within each scenario. Although the underlying random variable \(X\) is the same, a general law-invariant functional need not admit such an additive decomposition linking its overall variability to these scenario-based components. This motivates a fundamental question: which law-invariant functionals satisfy the decomposition principle?
 
%  The decomposition principle \eqref{eq:DP} states that the value $\rho(X)$ can be written as the sum of the between-group component $\rho\bigl(\E[X\mid\mathcal G]\bigr)$ and the 
% expected within-group component $\E[\rho(X\mid\mathcal G)]$. This can be seen as a generalization of the law of total variance, which is the case
% when $\rho$ is the variance. A natural question is whether the decomposition
% principle, possibly together with other conditions, can characterize the Jensen-gap functionals. 

% The preceding examples motivate the following inverse problem:
% under what conditions must a functional of random variables have the form
% \begin{equation}\label{eq:rho_jensen_form_intro}
%     \rho(X)
%     =
%     \E[f(X)]-f(\E[X])
% \end{equation}
% for some generator $f$? Equivalently, can the Jensen-gap representation be
% derived from properties of the functional $\rho$ itself?

The closest work to the above problem is \cite{campbell2025excessgrowth}, who provided several characterizations of the excess growth rate functional, which corresponds to $f=-\log$. 
We provide a concise result to complete the picture:
 on an atomless probability
space, binary
decomposition (BD) --- a weaker version of \eqref{eq:DP} --- and law invariance (LI) jointly  characterize
the Jensen-gap representation on simple random variables. Under suitable nonnegativity and continuity
conditions, we extend the results to broader classes of random
variables, including integrable ones.  

We then study how additional conditions identify the generator \(f\). When comparing financial positions with the same expected payoff, it is natural to ask whether their relative variability should depend on investment scale or the level of wealth. In particular, one may require that scaling both positions proportionally (\(X\mapsto cX\)) or adding the same deterministic cash amount (\(X\mapsto X+c\)) should not reverse their variability ordering. These considerations motivate scaling invariance (SI) and translation invariance (TI), respectively. We show that (SI) implies homogeneity of some degree \(\alpha\), a property of variability measures \citep[Definition~1]{bellini2022parametric}, and leads to logarithmic, \(x\log x\), and power generators. Alternatively, (TI) leads to exponential and quadratic generators. In particular, the logarithmic Jensen gap and variance are special cases. 
%We also show that convexity of nonnegative Jensen-gap functionals characterizes variance up to a nonnegative multiplicative constant. %A related characterization can be deduced from \citet[Theorem~4.4]{Chafai2006} under smoothness assumptions on the generator, whereas our result does not require such assumptions.

% We then study how additional conditions identify the
% generator $f$. In the context of variability measures, it is often desirable that applying the same scaling ($X\mapsto cX$) or translation ($X\mapsto X+c$) to
% two random variables should not reverse their ordering under $\rho$. Motivated by
% this idea, we consider two alternative conditions to identify the generator
% $f$. Scaling invariance (SI) requires that a common scaling preserves the
% ordering under $\rho$. We show that (SI) implies homogeneity of some degree
% $\alpha$, which is one property of the variability measures \citep[Definition~1]{bellini2022parametric}. This includes logarithmic, $x\log x$, and power generators.
% Alternatively, translation invariance (TI) requires that a common
% translation preserves the ordering under $\rho$, and leads to exponential
% and quadratic generators. In particular, the logarithmic Jensen gap and
% the variance are special cases.  

Our paper is organized as follows. Section~\ref{s:jensen_gap} introduces the Jensen gaps and discusses their functional properties, including the decomposition principle. 
In particular, we obtain a unified equivalence result on 
many properties that   characterize the convexity of the generator, as well as a simple characterization of variance among this class. 
Since the decomposition principle relies
on conditional functionals, Section~\ref{s:con_finite} introduces the formal
definitions of conditional functionals with respect to a finite
sub-$\sigma$-algebra $\mG$. In Section~\ref{ss:chara_jen_gap}, we
show that law invariance and binary decomposition  characterize the
Jensen-gap representation on simple random variables, and then extend this
representation to larger classes of random variables, including compactly supported ones and integrable
ones, under suitable additional conditions. In Section~\ref{s:identi_f},
starting from the Jensen-gap representation, we identify the generator $f$
through two alternative routes based on scaling invariance (SI) and
translation invariance (TI).
% We also compare our results with three characterization theorems of the excess growth rate in
% \citet{campbell2025excessgrowth}. 
Section~\ref{s:cond_fun_general_G} extends the conditional framework from finite sub-$\sigma$-algebras to general ones, and shows how the results developed under finite conditioning structures can be extended to this more general setting. Section~\ref{s:conclu} concludes the paper.

\textbf{Related technical results in the literature.}
\citet{ebanks2008jensen} characterizes differences of the form
$f(x)+f(y)-\lambda f(\mu(x+y))$, where $\lambda$ and $\mu$ are given
parameters. Under $\lambda=2$ and $\mu=1/2$, this difference becomes two times the Jensen gap of a random variable, which takes values $x$ and $y$ equiprobably. \citet{olbrys2021taffine} characterizes $t$-affine differences
of the form
\(
t f(x)+(1-t)f(y)-f\bigl(tx+(1-t)y\bigr)
\)
for a fixed $t\in(0,1)$, while \citet{olbrys2022affine} treats the
corresponding affine difference with $t$ also varying. These results
provide characterizations of two-point Jensen gaps and are
closely related to the binary structure in our analysis.
% They only concern the two-point case, corresponding to Jensen
% gaps of two-valued simple random variables.
% \citet{ebanks2008jensen} characterizes the differences of the form $f(x)+f(y)-\lambda f(\mu(x+y))$ where $\lambda$ and $\mu$ are given parameters.
% \citet{olbrys2021taffine} and \citet{olbrys2022affine} characterize maps
% that can be represented as affine or $t$-affine differences of the form
% $
%     t f(x)+(1-t)f(y)
%     -
%     f\bigl(tx+(1-t)y\bigr).
% $
% These results provide characterizations of Jensen-type differences and
% are closely related to the two-point structure in our analysis.
% However, they only concern the binary case, corresponding to the Jensen gaps
% of two-valued simple random variables. 
\citet{campbell2025excessgrowth} provide three theorems to characterize the excess growth rate, corresponding to the logarithmic
generator. Most relevant to our work, their Theorem~3.1 introduces a
chain rule (condition (A7) in their paper), which is similar to our decomposition principle (see (BD) of our paper). \citet{campbell2025excessgrowth} use this chain rule and some other conditions to characterize the logarithmic Jensen gap through
relative entropy. They do not reveal that the decomposition principle, together
with other suitable conditions, can characterize the general class of
Jensen-gap functionals beyond the logarithmic case. Their Theorem~3.11  treats general $f$, but an affine condition (D4) already assumes a  form that is close to the Jensen gap.

The Jensen gap with a convex generator satisfies many properties that are typically assumed for deviation measures \citep{rockafellar2006generalized,grechuk2009maximum} and variability measures \citep{furman2017gini, bellini2022parametric}, such as $\rho(c)=0$ for constants $c$, nonnegativity, mixture-concavity, and convex-order consistency (see Section \ref{s:jensen_gap}). However,  up to scaling, this class only contains one convex variability measure on $L^\infty$, that is, the variance (Proposition~\ref{prop:JG-convex}). 

As explained above, our construction of $\rho(X\mid \mathcal G)$ is closely related to conditional risk measures. Different from the majority of the   literature on conditional risk measures,  we do not assume convexity of the functional.
 A decomposition identity related to (BD) is used by \citet{PistoriusStadje2017} as a time consistency condition for a family of conditional deviation measures that are convex. 
We refer to \cite{ruszczynski2010_risk_averse_dp}, \cite{hoffmann2016risk}, and  \cite{pichler2017_quantitative} for some more developments on conditional risk measures.  

 \textbf{Notation.} Throughout this paper, we work on an atomless probability space $(\Omega,\mathcal F,\p)$. 
 Almost surely equal random variables are treated as identical. 
As usual, we denote 
by $L^q$, $q>0$,  the space of all random variables $X$ with a finite $q$-th moment, by $L^\infty$   the space  of essentially bounded random variables, and by $\mathcal X=L^0$ the space  of all  random variables. 
A random variable is \emph{simple} if it takes finitely many values.   The equality $X\laweq Y$ means that $X$ and $Y$ have the same distribution. The quantile function of a random variable $X$ at level $\alpha\in(0,1)$ is $Q_X(\alpha)=\inf\{t\in\R:\p(X\leq t)\geq\alpha\}$. Let $\RR=[-\infty,\infty]$, $\R_+=(0,\infty)$, $a\vee b=\max(a,b)$, $a\wedge b=\min(a,b)$, $a^+=\max(a,0)$, and $a^-=\max(-a,0)$. 
 Moreover, $\#A$ means the number of elements in the set $A$. Measurability of real-valued functions means Borel measurability. The terms “increasing” and “decreasing” are understood in the non-strict sense. Whenever at least one of \(X^+\) and \(X^-\) is integrable, both \(\mathbb E[X]\) and \(\mathbb E[X\mid\mathcal G]\) are well-defined in the extended sense and may take the value \(+\infty\) or \(-\infty\) \citep[Remark~8.16]{klenke2020probability}.
%  Whenever at least one of $X^+$ and $X^-$ is integrable,
% the expectation $\E[X]$ and the conditional expectation $\E[X\mid\mG]$ are well-defined. And they are understood in the standard
% extended sense and are allowed to take the value $\infty$ or $-\infty$ \citep[Remark~8.16]{klenke2020probability}.
We adopt the convention $0\times \infty=0$.
\section{Jensen gaps and their   properties}\label{s:jensen_gap}
In this section, we define Jensen gaps, discuss  
  their properties, 
  present characterization results on various forms of convexity, and present the decomposition
principle. 

\subsection{Definition of Jensen gaps}

We first formally define the main object of the paper.
\begin{definition}\label{def:jensen_gap}
A \emph{Jensen gap} is a functional
$J_f:\X_J\to\RR$, where
$\X_J$ is any subset of $L^1$, of the form
\begin{equation}
\label{eq:def-JG}        J_f(X)=\E[f(X)]-f(\E[X])
    \quad \mbox{ 
for all $X\in \mathcal X_J$} 
\end{equation}
for some real-valued function $f$, where the right-hand
side is well-defined for every $X\in\X_J$.\footnote{For \eqref{eq:def-JG} to be well-defined, four conditions
are required: the domain of $f$ contains the range of $X$,
$\E[X]$ is well-defined and belongs to the domain of $f$,
$\E[f(X)]$ is well-defined, and the expression does not
involve $\infty-\infty$.} 
The function $f$ is called a \emph{generator}, and
$J_f$ is also called an \emph{$f$-Jensen gap}.
\end{definition}

When $X\in \X_J$ is simple,  $f(X)$ is also simple regardless the measurability of $f$. Thus,  we do not impose the measurability of $f$ in general. When elements of $\X_J$ take continuous values in an interval,
measurability of $f$ on that interval is needed for $J_f$ in \eqref{eq:def-JG} to be well-defined.  
 If $f:\R\to \R$ is convex, then   $ \E[f(X)]-f(\E[X])$  in \eqref{eq:def-JG} 
is well-defined for all   $X\in L^1$. Therefore, $J_f$ for a convex $f$ can always be defined on $L^1$.

Several elementary properties follow directly from
the definition. A Jensen gap is law-invariant (i.e., $J_f(X)=J_f(Y)$ when $X\laweq Y$) and
vanishes on constants. Moreover, adding an affine
function to the generator does not change the
Jensen gap. If $f$ is convex, then $J_f\geq0$ by Jensen's inequality.

\subsection{Various forms of convexity}
Four inter-connected notions 
are important for risk measures and variability measures, defined below. 
Let $\rho:\X_\rho\to\RR$ be a 
law-invariant mapping i.e., $\rho(X)=\rho(Y)$ whenever $X\laweq Y$, and we assume  that $\X_\rho$ is a convex subset of $L^1$.
Write $\mu_X$ for the law of $X$ and define
$\tilde\rho(\mu_X)=\rho(X)$. All summations below are assumed to be well-defined.
\begin{enumerate}[label=(\alph*)]
    \item Convexity:
 $
    \rho(tX+(1-t)Y)\le t\rho(X)+(1-t)\rho(Y)
 $
 for all $t\in (0,1)$ and $X,Y\in\X_\rho$.
    \item Quasi-convexity:
 $
    \rho(tX+(1-t)Y)\le\max\{\rho(X),\rho(Y)\}
 $   for all $t\in (0,1)$ and $X,Y\in\X_\rho$.
    \item Mixture concavity: the law domain
    $ \{\mu_X:X\in\X_\rho\}$ is convex and $    \tilde\rho$ is concave. 
    \item Convex-order consistency (also known as Schur-convexity):
$X\le_{\mathrm{cx}}Y $ implies $ \rho(X)\le\rho(Y), $
    where  $X\le_{\mathrm{cx}}Y$ means $\E[\phi(X)]\le\E[\phi(Y)]$ for every convex function
    $\phi:\R\to\R$.
\end{enumerate}

To interpret these properties, 
the usual convexity reflects the consistency with respect to diversification, and it  is standard for risk measures \citep{FS02}  and  deviation measures  \citep{rockafellar2006generalized}.
Further, \cite{CMMM11} argued that quasi-convexity is a better notion for diversification  when cash additivity is not imposed. 
Mixture concavity 
has the interpretation that external randomization increases riskiness or variability, because the mixture operation on distributions represents random sampling from these distributions; see \cite{acciaio2013law}. 
  {Convex-order consistency}
reflects that $\rho$ is consistent with a mean-preserving spread; 
see \cite{bauerle2006stochastic} and \cite{mao2020risk} for characterizations of convex-order consistency for risk measures. 

The four properties above 
are equivalent for the class of 
law-invariant signed Choquet integrals \citep[Theorem 3]{wang2020characterization}, including many risk measures and variability measures,  but  these equivalences do not extend to Jensen gaps.
 The next proposition summarizes these properties for Jensen gaps in a unified result. 

 \begin{proposition}
 \label{prop:JG-convex}
Let $q\in [1,\infty]$. For the $f$-Jensen gap $J_f:L^q \to \overline{\R}$  with a measurable generator $f:\R\to\R$, the  following are equivalent:
 \begin{enumerate}[label=(\roman*)]
  \item $f$ is convex; 
   \item $J_f $ is nonnegative;

 \item  $J_f$ is mixture-concave;
 \item  $J_f$ is convex-order consistent;
 \item $J_f$ is convex on the section $\{X\in L^q:\E[X]=c\}$ for every $c\in \R$;
 \item $J_f$ is convex on the section $\{X\in L^q:\E[X]=c\}$ for some $c\in \R$\blue{;}
 \item   $J_f$ is quasi-convex on the section $\{X\in L^q:\E[X]=c\}$ for every $c\in\R$;
 \item   $J_f$ is quasi-convex on the section $\{X\in L^q:\E[X]=c\}$ for some $c\in\R$.
 \end{enumerate}
Furthermore, the  following are equivalent:
 \begin{enumerate}[resume*]
  \item  $J_f=a \var$ for some $a\ge 0$, where we set $\var=\infty$ outside $L^2$;
 \item   $J_f$ is convex;
 \item   $J_f$ is quasi-convex.
 \end{enumerate}
 \end{proposition}

Some of the equivalences in Proposition~\ref{prop:JG-convex} are straightforward or  follow  directly from  the literature, and we provide a unified and self-contained proof in \ref{app:proof-2}. 
Another natural weakening of convexity  for risk measures is star-shapedness, reflecting   liquidity concerns when risk exposure increases  \citep{CCMTW22}. A characterization of star-shaped Jensen gaps is given in  \ref{app:JG-star}, where we see that star-shapedness is not equivalent to  any conditions above.

\begin{remark}\label{re:chafai_cx}
When $\X_J$ is smaller than $L^\infty$, 
convexity  may not force $J_f$ to be a multiple of variance, as studied by \cite{Chafai2004,Chafai2006}.
In particular, 
\citet[Theorem~4.4]{Chafai2006} shows that,
for a smooth generator $f$ on an open interval $I$
and $\mathcal X_J$ being the set of random variables taking values in compact
subintervals of $I$, the Jensen gap is convex
if and only if $f$ is affine, or $f''>0$
and $1/f''$ is concave on $I$. Some common examples are $f(x)=x^2$ for $I=\R$, $f(x)=x\log x$ for $I=\R_+$ and $f(x)=x^p$ for $I=\R_+$ and $p\in(1,2]$.
When $I=\R$, the positive concave
function $1/f''$ must be constant, so this gives the case $J_f=a\var$
for some $a\geq0$.  Our equivalence result in Proposition~\ref{prop:JG-convex}   does not need any
smoothness assumptions on $f$ and imposes the weaker quasi-convexity.
\end{remark}

Closely related to the convex order, 
the 
\emph{dilation order} \(X\leq_{\mathrm{dil}}Y\) is defined as   \(X-\E[X]\leq_{\mathrm{cx}}Y-\E[Y]\).
We record a characterization of  $\le_{\rm cx}$ and $\le_{\rm dil}$ by    Jensen gaps in the next simple result.
\begin{proposition}
\label{prop:convex-order} 
 For any $X,Y\in  L^1$, we have
\begin{align*}
  X\le_{\rm cx } Y ~& \iff~ \E[X]=\E[Y] \mbox{ and } J_f(X)\le J_f(Y) \mbox{ for all convex $f:\R\to\R$};
\\
  X \le_{\rm dil } Y  ~& \iff~    J_f(X-\E[X])\le J_f(Y-\E[Y]) \mbox{ for all convex $f:\R\to\R$},
\end{align*}
 where $J_f$ is the $f$-Jensen gap defined on $L^1$.
\end{proposition}
  Proposition~\ref{prop:convex-order} can be easily shown: 
If $X\le_{\mathrm{cx}}Y$, testing with $f(x)=x$ and $f(x)=-x$
gives $\E[X]=\E[Y]$. Whenever $\E[X]=\E[Y]=m$,
 $
J_f(X)\le J_f(Y)
 \Longleftrightarrow 
\E[f(X)]\le\E[f(Y)],
$ 
since the common finite term $f(m)$ cancels.
This proves the first equivalence.
Applying it to $X-\E[X]$ and $Y-\E[Y]$, whose means are both
zero, proves the second by the definition of dilation order.

 Both the convex order and the dilation order are  known as variability orders; 
see \citet[Section 3.A.1]{shaked2007stochastic} for a comprehensive treatment, and therefore 
Proposition~\ref{prop:convex-order}  strengthens the interpretation of Jensen gaps as  measures of variability.

\subsection{The decomposition principle}

We now turn to 
 the
\emph{decomposition principle}, which will be the central object  in our later sections.
For a sub-$\sigma$-algebra $\mG\subseteq\mathcal F$,
the conditional Jensen gap is defined as
\(
    J_f(X\mid\mG)
    =\E[f(X)\mid\mG]-f(\E[X\mid\mG]).
\)
By the tower property of conditional expectation, we obtain  
 $$
    J_f(X)
    =
    \E[J_f(X\mid\mG)]
    +J_f(\E[X\mid\mG]),
 $$
whenever all terms are well-defined.
% This identity decomposes the Jensen gap into
% the expected conditional Jensen gap $\E[J_f(X\mid\G)]$ and the Jensen gap
% of the conditional expectation $J_f(\E[X\mid\G])$. 
Conversely, in Section~\ref{s:jensen_func}, we will show that a law-invariant functional satisfying the decomposition principle must be a Jensen gap on simple random variables, with extensions to compactly supported ones and integrable ones under suitable continuity conditions. However, in order to specify the decomposition principle for a general functional $\rho$, we need to understand what $\rho(X\mid \mathcal G)$ means, which is not a familiar object for a general functional $\rho$.  For this, we introduce conditional law-invariant functionals in the next section,  which include conditional law-invariant risk measures in finance as special cases. 
\section{Conditional risk measures on a finite sigma-algebra}\label{s:con_finite}
% In this section, we illustrate the definition of the conditional version of the law-invariant risk measures, which are conditional on the finite sub-$\sigma$-algebra $\mathcal G$. Here, we further assume that $\mathcal G=\sigma(A_1,\ldots,A_n)$, where $A_1,\ldots,A_n$ form a finite measurable partition of $\Omega$ with $\mathbb P(A_i)>0$ for every $i$. Then we show that the conditional version of $\E [f(X)]$ is indeed $\E[f(X)\mid\mathcal G]$, and the conditional version of the Expected Shortfall $\ES_\alpha(X)$ admits the form given in Eq.~\eqref{eq:con_ES_finite}. The case of general sub-$\sigma$-algebra is discussed in Section~\ref{s:cond_fun_general_G}.
In this section, we define conditional versions of law-invariant functionals with respect to a finite sub-$\sigma$-algebra $\mathcal G=\sigma(A_1,\ldots,A_n)$, where $A_1,\ldots,A_n$ form a finite measurable partition of $\Omega$.
 The construction applies to monetary risk measures, such as Expected Shortfall, as well as measures of variability, such as the variance. 
% with $\mathbb P(A_i)>0$ for all $i=1,...,n$. 
We show that the conditional version of $\E[f(X)]$ is $\E[f(X)\mid\mathcal G]$, and derive the conditional representation of Expected Shortfall in \eqref{eq:con_ES_finite}. The extension to general sub-$\sigma$-algebras is presented in Section~\ref{s:cond_fun_general_G}.
% \begin{definition}(Conditional law on finite $\sigma$-algebra)\label{def:con_law}
%    Let $\mu_X=\mathbb P\circ X^{-1}$ denotes the distribution of an $S$-valued valued random variable $X$, where $S\subseteq\R$. And let $\mathcal G\subseteq\mathcal F$ be a finite sub-$\sigma$-algebra. Let
% $\mu_{X\mid\mathcal G}$ be a regular conditional distribution of $X$ given
% $\mathcal G$, that is, for every Borel set $B$
% \[
% \mu_{X\mid\mathcal G}(\omega,B)
% =
% \mathbb P(X\in B|\mathcal G)(\omega)=\mathbb E[\id_{\{X\in B\}}|G](\omega)\quad\text{a.s.},
% \]
% and $\mu_{X\mid\mathcal G}(\cdot,B)$ is $\mathcal G$-measurable and $\mu_{X\mid\mathcal G}(\omega,\cdot)$ is a probability measure for every $\omega\in\Omega$.
% \end{definition}
\begin{definition}
\label{def:con_law}
Let $X$ be an $S$-valued random variable, where
$S\subseteq\mathbb R$ is a Borel set, and let $\mathcal B(S)$ denote the Borel $\sigma$-algebra on $S$. Let
\(
    \mu_X=\p\circ X^{-1}
\)
denote its distribution. Let $\mathcal G\subseteq\mathcal F$ be a finite
sub-$\sigma$-algebra. A conditional law of $X$ given $\mathcal G$ is denoted
by $\mu_{X\mid\mathcal G}$ and satisfies, for every
$B\in\mathcal B(S)$,
\begin{equation}\label{eq:con_law_def}
        \mu_{X\mid\mathcal G}(\cdot,B)
    =
    \p(X\in B\mid\mathcal G)
    =
    \E\!\left[
        \id_{\{X\in B\}}
        \mid \mathcal G
    \right]
    \quad\text{a.s.}
\end{equation}
Moreover, $\mu_{X\mid\mathcal G}(\cdot,B)$ is
$\mathcal G$-measurable for every
$B\in\mathcal B(S)$, and
$\mu_{X\mid\mathcal G}(\omega,\cdot)$ is a probability measure on
$\mathcal B(S)$ for every $\omega\in\Omega$.
\end{definition}
% Since $\mathcal B(S)$ is countably generated, the exceptional null set
% may be chosen independently of $B$; that is, there exists
% $N\in\mathcal G$ with $\mathbb P(N)=0$ such that the above identities in Eq.~\eqref{eq:con_law_def}
% hold for every $\omega\notin N$ and every $B\in\mathcal B(S)$ \citep[Lemma~10.4.3]{bogachev2007}. 
If $\mathcal G=\sigma(A_1,\ldots,A_n)$, where
$A_1,\ldots,A_n$ form a finite measurable partition of $\Omega$, let
\(
I_+=\{i\in\{1,\ldots,n\}:\p(A_i)>0\}.
\)
For every $i\in I_+$, define
\[
\mu_{X\mid A_i}(B)
=
\p(X\in B\mid A_i)
=
\frac{\p(A_i\cap\{X\in B\})}{\p(A_i)},
\qquad B\in\mathcal B(S).
\]
Then, for every $B\in\mathcal B(S)$,
\begin{align*}
\mu_{X\mid\mathcal G}(\omega,B)
&=
\E\!\left[
\id_{\{X\in B\}}
\mid\mathcal G
\right](\omega)=
\sum_{i\in I_+}
\E\!\left[
\id_{\{X\in B\}}
\mid A_i
\right]
\id_{A_i}(\omega)=
\sum_{i\in I_+}
\mu_{X\mid A_i}(B)\id_{A_i}(\omega)
\quad\text{a.s.}
\end{align*}
Thus we can simply choose the version $\mu_{X\mid\mathcal G}(\omega,\cdot)=\sum_{i\in I_+}
\mu_{X\mid A_i}(\cdot)\id_{A_i}(\omega)$ for all $\omega\in \bigcup_{i\in I_+} A_i$.

% And if $\p(A_i)>0$ for $i=1,..,n$, we can choose $\mu_{X\mid\G}(w,\cdot)=\sum_{i=1}^n
% \mu_{X\mid A_i}(B)\id_{A_i}(\omega)$.
% If $\mathcal G=\sigma(A_1,\ldots,A_n)$, where $A_1,\ldots,A_n$ form a finite measurable partition of $\Omega$ with $\mathbb P(A_i)>0$ for every $i$, we have \begin{align*}
% \mu_{X\mid\mathcal G}(\omega,B)&=\mathbb E[\id_{\{X\in B\}}\mid\mathcal G](\omega)
% =
% \sum_{i=1}^n
% \mathbb E[\id_{\{X\in B\}}\mid A_i]\id_{A_i}(\omega)=
% \sum_{i=1}^n
% \mu_{X\mid A_i}(B)\id_{A_i}(\omega)
% \quad\text{a.s.,}
% \end{align*}
% where $$
% \mu_{X\mid A_i}(B)
% =
% \mathbb P(X\in B\mid A_i)
% =
% \frac{\mathbb P(A_i\cap\{X\in B\})}{\mathbb P(A_i)}.
% $$
% Thus, if $\p(A_i)>0$ for every $i$, we can choose the following version of $\mu_{X\mid\mG}$:
% \begin{equation}\label{eq:con_mu_G_finite}
%     \mu_{X\mid\mathcal G}(\omega,\cdot)
% =
% \sum_{i=1}^n
% \id_{A_i}(\omega)\mu_{X\mid A_i}(\cdot).
% \end{equation}

Recall that a functional $\nu:\mathcal X_\nu\to  \RR $ is law-invariant if
$X\laweq Y$ implies $\nu(X)=\nu(Y)$ for all
$X,Y\in\mathcal X_\nu$, where
$\mathcal X_\nu\subseteq\mathcal X$. In this case, there exists a functional $\tilde{\nu}$ such that $\nu(X)=\tilde{\nu}(\mu_X)$ for all $X$. Note that the conditional law is defined only up to a.s.~equality.
Therefore, in the following, if it takes values in the corresponding domain a.s.,
we may choose a version that does so everywhere. This leads to the following Definition~\ref{def:con_risk_meas_fini} of the conditional functional on the finite $\sigma$-algebra. 
    % \begin{equation*}
    %     \nu(X)=\tilde{\nu}(\mu_X)
    % \end{equation*}
\begin{definition}\label{def:con_risk_meas_fini}
    For a law-invariant functional $\nu$, whenever
$\mu_{X\mid\mG}$ belongs to the domain of $\tilde\nu$ a.s., we define its conditional version as 
    $
        \nu(X\mid\mG)=\tilde{\nu}(\mu_{X\mid\mG}).
   $
   Since $\mathcal G=\sigma(A_1,\ldots,A_n)$, we have
    $
\nu(X\mid\mathcal G)
=
\sum_{i\in I_+} \id_{A_i}\nu(X\mid A_i)
$ a.s.,
where
$
\nu(X\mid A_i)
=
\tilde{\nu}(\mu_{X\mid A_i}).
$ 
% And if $\p(A_i)>0$ for $i=1,..,n$, $\nu(X\mid\G)=\sum_{i=1}^n
% \nu(X\mid A_i)$.
\end{definition}
Based on the above definitions, we can derive the conditional versions of two common functionals:
$\E[f(X)]$ and the Expected Shortfall $\ES_\alpha(X)$ in the following examples.
\begin{example}\label{ex:con_E_finite}
   Let $\nu(X)=\E[f(X)]=\int f(x)\mu_X(\d x)$, where $f$ is measurable, and at least one of $f^+(X)$ and $f^-(X)$ is integrable. From Definition~\ref{def:con_risk_meas_fini}, its conditional version is $\nu(X\mid\mG)=\sum_{i\in I_+} \id_{A_i} \int f(x)\mu_{X\mid A_i}(\d x)=\sum_{i\in I_+} \id_{A_i} \E[f(X)\mid A_i]$ a.s. As $\E[f(X)\mid\mG]=\sum_{i\in I_+} \id_{A_i} \E[f(X)\mid A_i]$ a.s., we obtain $\nu(X\mid\mG)=\E[f(X)\mid\mG]$ a.s.
\end{example}
\begin{example}\label{ex:con_E_X_Y_finite}
Example~\ref{ex:con_E_finite} can be extended as follows. We assume that the relevant expectations in the following are well-defined. If $Y=\sum_{i\in I_+} \id_{A_i} y_i$ a.s., then we have that the conditional version of $\E[f(X,Y)]$ on $\mG=\sigma(A_1,...,A_n)$ is $\sum_{i\in I_+} \id_{A_i} \E[f(X,y_i)\mid A_i]$. Indeed, let $Z=f(X,Y)$ and apply Example~\ref{ex:con_E_finite} to the expectation
functional $\nu(W)=\E[W]$, then we obtain that $\nu(Z\mid \mG)=\sum_{i\in I_+} \id_{A_i} \E[Z\mid A_i]$ a.s. As $\E[Z\mid A_i]=\E[\id_{A_i} Z]/\p(A_i)$ and $\id_{A_i}Z=\id_{A_i}f(X,y_i)$ a.s. for $i\in I_+$, we obtain that $\nu(Z\mid\mG)=\sum_{i\in I_+} \id_{A_i} \E[f(X,y_i)\mid A_i]$ a.s., which equals $\E[f(X,Y)\mid \mG]$ a.s.
\end{example}
Based on the above results, we can derive the conditional version of the Expected Shortfall $\ES_\alpha(X)$ in the following Example~\ref{ex:con_ES_G_finite}. For $\alpha\in(0,1)$ and an integrable loss random variable $X$, the Expected Shortfall has the variational representation \citep{rockafellar_uryasev_2002}
\begin{equation*}
    \operatorname{ES}_\alpha(X)=
\min_{m\in\mathbb R}
\left\{
m+\frac{1}{1-\alpha}\E\left[(X-m)^+\right]
\right\}.
\end{equation*}
\begin{example}\label{ex:con_ES_G_finite}
 Let $\alpha\in(0,1)$ and $X\in L^1$. From Definition~\ref*{def:con_risk_meas_fini}, the conditional Expected Shortfall satisfies 
\begin{equation}\label{eq:con_ES_finite}
    \ES_\alpha(X\mid \mG)=\essinf_{Z\in L^0(\mG)}\left\{Z+\frac1{1-\alpha}\E[(X-Z)^+\mid \mG]\right\}\quad\text{a.s.,}
\end{equation}
which is the standard form given in  \citet[Example~6.2]{RuszczynskiShapiro2006}.
 A self-contained derivation is given in \ref{app:proof_gener_G}.
\end{example}
\section{Characterization of Jensen-gap functionals}\label{s:jensen_func}
% In this section, we first characterize Jensen-gap functionals in
% Section~\ref{ss:chara_jen_gap}. We then turn to the identification of
% the generator $f$ in Section~\ref{s:identi_f}, where additional
% properties of $\rho$ are used to determine its possible forms within the
% Jensen-gap class. 
In this section, we first characterize Jensen-gap functionals on simple
random variables under law invariance and binary decomposition. We then
extend the representation to $\mathbb D$-compactly supported random variables and integrable random variables under
suitable conditions.
We use the following notation with a set $\mathbb D\subseteq\R$.
% $\mathbb D$ denotes an interval and $\tilde{\mathbb D}$ denotes an open interval.
\begin{enumerate}[label=(\roman*)]
    \item Let $\mathcal{X}_s(\mathbb D)$ be the set of simple random variables taking values in $\mathbb D$.
    \item When $\mathbb D$ contains an interval, 
we say that a random variable $X$ is \emph{$\mathbb D$-compactly supported} if $X$ takes values in $\mathbb D$ and $\p(X\in [a,b])=1$ for some $[a,b]\subseteq \mathbb D$. 
\item 
Let 
$\mathcal X_c( {\mathbb D})$ be the set of  ${\mathbb D}$-compactly supported random variables.
    \item Let $\mathcal X_1(\mathbb D)$ be the set of integrable random variables taking values in $\mathbb D$. 
%     \item $\mathcal X_e(\mathcal D)$ is the set of equiprobable simple
% random variables of the form
% $
% X=\sum_{i=1}^{2^n}x_i\mathbf 1_{A_i},
% $ where $x_i\in\mathcal D$, $A_1,...,A_{2^n}$ form a measurable partition of $\Omega$, $\mathbb P(A_i)=2^{-n}$ and $i=1,\ldots,2^n$ for $n\geq1$. The case of $n=0$ refers to the constant $X$, which is also included in $\mathcal X_e(\mathcal D)$.
\end{enumerate}
%\subsection{Characterization of Jensen-gap functionals}
\label{ss:chara_jen_gap}
We will use the following properties of a functional $\rho:\X_\rho\to \RR$ in our main characterization results. The following properties are stated for all random variables or constants  in $\X_\rho$. 
\begin{enumerate} 
    \item[(LI)] Law-invariance: If $X \overset{\text{d}}{=} Y$, then $\rho(X)=\rho(Y)$.
    \item[(C)] Constancy: $\rho(c)=0$ for any constant $c$. 
    % \textcolor{red}{this can be deleted or add some comments, since from (BD), let X=c, $\rho(c|G)=\rho(c)$ a.s., so $\rho(c)=\rho(c)+\rho(c)=0$, we have $\rho(c)=0$ }
    \item[(BD)] Binary decomposition: $\rho(X)=\E[\rho(X\mid\mathcal G)]+\rho(\E[X\mid\mathcal G])$ for every binary $\sigma$-algebra $\mathcal G\subseteq \mathcal F$, that is, $\mG=\sigma(A,A^c)=\sigma(A)$ for some measurable set $A$.
    \item[(N)] Nonnegativity: $\rho(X)\geq 0$.
\end{enumerate}
Note that if $\rho(c)\in\R$, then (LI) and (BD) imply (C) by taking $X=c$. 

The main condition is the decomposition principle (BD), which has a natural probabilistic interpretation.
It decomposes the overall value into a between-group component
$\rho(\E[X\mid\mG])$ and an expected within-group component
$\E[\rho(X\mid\mG)]$. In particular, when $\rho$ is the variance, (BD) reduces to
the law of total variance
\(
\var(X)
=
\E[\var(X\mid\mG)]
+
\var(\E[X\mid\mG]).
\) More generally, Jensen-gap functionals naturally satisfy the decomposition
principle for any finite sub-$\sigma$-algebra $\mathcal G$. However, we show that,
to characterize the Jensen-gap functionals, it is sufficient to impose
this principle only for binary sub-$\sigma$-algebras, namely (BD). This
leads to the following theorem.
\begin{theorem}\label{thm:rho_form_simple}
For $\rho:\mathcal X_s(\mathbb D)\to\mathbb R$
with an interval $\mathbb D$, the following statements are equivalent:
\begin{enumerate}[label=(\roman*)]
\item $\rho$ satisfies (LI) and (BD);
\item  $\rho$ is a Jensen gap on $\mathcal X_s(\mathbb D)$.
% \item there exists a function $f:\mathbb D\to\mathbb R$
% such that $\rho$ is the Jensen gap of $f$, that is,
% \begin{equation}\label{eq:rho_form1}
%     \rho(X)=\E[f(X)]-f(\E[X]) \quad \mbox{
% for all 
% $X\in\mathcal X_s(\mathbb D)$.}
% \end{equation}
\end{enumerate} 
\end{theorem}
Note that  Theorem~\ref{thm:rho_form_simple}  does not require that $f$ is measurable, because $\rho$ is only defined on simple random variables. The proof of Theorem~\ref{thm:rho_form_simple} is given in \ref{app:proof_chara_jens}. Theorem~\ref{thm:rho_form_simple} characterizes the general Jensen-gap
representation, where the generator $f$ is not required to satisfy any
regularity or shape restriction. A natural additional requirement is
nonnegativity,
\(
\rho(X)\geq 0,
\)
especially when $\rho$ is interpreted as a measure of variability. For a Jensen-gap functional $\rho$, the nonnegativity of $\rho$ is equivalent to convexity
of $f$ by Jensen's inequality. Thus, within the class characterized in
Theorem~\ref{thm:rho_form_simple}, nonnegativity is exactly the condition
that identifies convex generators. We summarize this result in Theorem~\ref{thm:rho_form_nonne_convex}.
\begin{theorem}\label{thm:rho_form_nonne_convex}
For $\rho:\mathcal X_s(\mathbb D)\to\mathbb R$
with an interval $\mathbb D$, the following statements are equivalent:
\begin{enumerate}[label=(\roman*)]
\item $\rho$ satisfies (LI), (BD) and (N);
\item $\rho$ is an $f$-Jensen gap on $\mathcal X_s(\mathbb D)$ for some convex function $f:\mathbb D\to\mathbb R$.
\end{enumerate} 
\end{theorem}
% Suppose we let the random variables taking values on an open interval $\tilde{\mathbb D}$, and add two more properties on $\rho$ as follows.
% \begin{enumerate} 
%     \item[(N)] Nonnegativity: $\rho(X)\geq 0$.
%     \item[(BC)] Bounded continuity: Let $[a,b]\subseteq\tilde{\mathbb D}$. If $a\le X_n\le b$ a.s. for every $n\in\mathbb{N}_+=\{1,2,3,...\}$ and $X_n\rightarrow X$ a.s., then $\lim_{n\rightarrow\infty}\rho(X_n)=\rho(X)$.
% \end{enumerate}
% Then we can have results similar to Theorem~\ref{thm:rho_form_simple}, where the domain of $\rho$ is changed to $\mathcal X_c(\tilde{\mathbb D})$, which is the set of random variables $X$ taking values on $\tilde{\mathbb D}$, and there exists $[a_X,b_X]\subseteq\tilde{\mathbb D}$ such that $a_X\le X\le b_X$ a.s. 
As can be seen from the proof of Proposition~\ref{prop:JG-convex}, condition (N) in Theorem~\ref{thm:rho_form_nonne_convex} can also be replaced by mixture concavity or convex-order consistency. Since only simple random variables are involved, no measurability assumption on \(f\) is needed here.

Having identified convex generators on the domain
$\mathcal X_s({\mathbb D})$, it is natural to ask whether the same
Jensen-gap representation can be extended to a larger class of random
variables.
We next consider $\mathbb D$-compactly supported random variables for an open interval $\mathbb D$.  Note that a convex function on an open
interval ${\mathbb D}$ is continuous and every
$X\in\mathcal X_c({\mathbb D})$ can be approximated by  simple random variables taking values in the same compact subinterval of \(\mathbb D\).  
For such random variables, their corresponding continuity for
$\rho$ on $\mathcal X_c({\mathbb D})$ is
\emph{compact continuity},
%: On the domain of $\rho$,  
%if $X_n\rightarrow X$ a.s.~and $X_1,X_2,\dots$ are $[a,b]$-compactly supported for some closed interval $[a,b] $, then $\lim_{n\rightarrow\infty}\rho(X_n)=\rho(X)$.
  a common property in decision theory for modeling non-expected-utility functionals
\citep{chew1995schur,chateauneuf2007sure,he2026diversification}. 
Here, for the characterization result, we only need   semicontinuity instead of the full continuity.
\begin{enumerate}
    \item[(CLS)] Compact lower semicontinuity: On the domain of $\rho$, if $X_n\rightarrow X$ a.s.~and $X_1,X_2,\dots$ are $[a,b]$-compactly supported for some interval $[a,b]\subseteq\mathbb D$, then $\liminf_{n\rightarrow\infty}\rho(X_n)\ge \rho(X)$. 
     % \item[(CUS)] Compact upper semicontinuity: On the domain of $\rho$, if $X_n\rightarrow X$ a.s.~and $X_1,X_2,\dots$ are $[a,b]$-compactly supported for some closed interval $[a,b]$, then $\limsup_{n\rightarrow\infty}\rho(X_n)\leq\rho(X)$. 
\end{enumerate}

Compact upper semicontinuity (CUS) is defined by replacing 
$\liminf_{n\rightarrow\infty}\rho(X_n)\ge \rho(X)$
by $\limsup_{n\rightarrow\infty}\rho(X_n)\leq\rho(X)$ in (CLS).
As usual, compact continuity holds when both (CUS) and (CLS) hold.

Under (LI), 
we can replace $X_n\rightarrow X$ a.s. by $X_n\xrightarrow{\d} X$ in (CLS) or (CUS) without losing its strength. When $\mathbb D=\R$,  condition (CLS) is equivalent to the Fatou property commonly used in
the risk-measure literature 
\citep{delbaen2002coherent,jouini2006law,FS16}. 

% Here we adopt the convention
% \(
%     X\succeq_\rho Y
%     \Longleftrightarrow
%     \rho(X)\leq\rho(Y)
% \).
% This is because, when $\rho$ is interpreted as a measure of variability, a smaller
% value represents lower dispersion and is therefore regarded as preferable.
% \begin{enumerate} 
%     \item[(BC)] Bounded continuity: Let $[a,b]\subseteq\tilde{\mathbb D}$. If $a\le X_n\le b$ a.s. for every $n\in\mathbb{N}_+=\{1,2,3,...\}$ and $X_n\rightarrow X$ a.s., then $\lim_{n\rightarrow\infty}\rho(X_n)=\rho(X)$.
% \end{enumerate}
Now, we can establish a characterization of the Jensen gap on $\mathcal X_c({\mathbb D})$ with an open interval $\mathbb D$ in the following Proposition~\ref{thm:conti}.
% Thus, by imposing the bounded
% continuity condition (BC) on $\rho$, the representation established on
% the simple random variables can be extended to
% $\mathcal X_c(\tilde{\mathbb D})$ by the bounded convergence theorem.
% This is stated in the next Proposition~\ref{thm:conti}.
\begin{proposition}\label{thm:conti}
    For $\rho:\mathcal X_c({\mathbb D})\to\mathbb R$ with an open interval $\mathbb D$, the following statements are equivalent:
\begin{enumerate}[label=(\roman*)]
\item $\rho$ satisfies (LI), (BD), (N) and (CLS);

\item $\rho$ satisfies (LI), (BD), (N) and (CUS);

\item $\rho$ is an $f$-Jensen gap on $\mathcal X_c(\mathbb D)$ for some convex function $f:\mathbb D\to\mathbb R$.
\end{enumerate} 
\end{proposition}

As a simple consequence, the functional $\rho$ in (i)--(iii) of Proposition~\ref{thm:conti} 
indeed satisfies compact continuity, although only semicontinuity is assumed in (i) and (ii).  
The proof of Proposition~\ref{thm:conti} is given in \ref{app:proof_chara_jens}. 
The proofs under (CLS) and (CUS) are different because these are
one-sided semicontinuity conditions in opposite directions, and neither
generally implies the other. Although (CUS) for $\rho$ is formally (CLS) for $-\rho$, this does not yield symmetry in our setting because the transformation $\rho\mapsto-\rho$ does not preserve nonnegativity (N).

Condition (CLS) or (CUS) allows us to extend the representation to
$\mathcal X_c({\mathbb D})$ with open intervals $\mathbb D$. To further extend it to the larger
domain $\mathcal X_1({\mathbb D})$ of integrable random variables,
(CLS) and (CUS) are no longer sufficient, since the approximating random variables
need not take values in a common compact subinterval of
${\mathbb D}$. A natural conjecture is that one may impose an $L^1$-type continuity
condition on $\rho$ to extend the Jensen-gap representation to integrable ones. For example, one may consider the
following natural continuity requirement: if
$X_n,X\in\mathcal X_1(\mathbb D)$ for an open interval $\mathbb D$,
$X_n\to X$ a.s. (or in $L^1$), and
$|X_n|\leq Y$ a.s. for some $Y\in\mathcal X_1(\mathbb D)$, then
$\rho(X_n)\to\rho(X)$.

However, such a condition is not satisfied by Jensen-gap functionals with
general convex generators. The difficulty comes from the term
$\E[f(X)]$: since only $L^1$-integrability of $X$ is assumed,
$f(X)$ need not be integrable, and $f$ need not be monotone. Hence,
standard convergence theorems cannot in general be applied to obtain the
desired extension. Indeed, the following counterexample shows that even if
$X_n,X\in\mathcal X_1(\mathbb D)$,
$X_n\to X$ both a.s. and in $L^1$,
$|X_n|\leq Y$ a.s. for some
$Y\in\mathcal X_1(\mathbb D)$ with $|\rho(Y)|<\infty$, we may still have
\(
\rho(X_n)\nrightarrow\rho(X).
\) 
Besides, this example also shows that the compact-support condition in $\mathcal X_c(\mathbb D)$ of Proposition~\ref{thm:conti} cannot in general be removed. Once
this condition is dropped, it is possible to have $X_n\to X$ a.s. but
\(
\rho(X_n)\nrightarrow\rho(X).
\) This example is given as follows.
\begin{example}\label{ex:counter_L1_comp}
    Let
\(
\rho(X)=\log(\E[X])-\E[\log X]\), \(\mathbb D=\R_+\) and $U\sim\operatorname{Unif}(0,1)$. Define
\(X_n=\exp\left(-1/(nU)\right)\in(0,1)\) and \(X=1\).
Then $X_n,X\in\mathcal X_1(\mathbb D)$, $X_n\to X$ a.s.~and in
$L^1$, and $|X_n|\leq 1$ a.s. Note that
\(
\E[\log X_n]
=- \E\left[1/U\right]/n
=-\infty,
\)
and hence $\rho(X_n)=\infty$ for every $n$, whereas $\rho(X)=0$.
Therefore,
\(
\rho(X_n)\nrightarrow\rho(X).
\)
\end{example}

Hence, we seek a different continuity condition for extending the
Jensen-gap representation to integrable random variables. It turns out that continuity with respect to truncation is sufficient for this purpose.

% Recall that
% adding an affine function to a convex generator does not change its
% Jensen gap. Therefore, for $c\in\mathbb D$, we may normalize the
% generator so that it attains its minimum at $c$. The resulting generator
% is decreasing on the left of $c$ and increasing on the right of
% $c$.

% This observation suggests considering sequences satisfying
% $
% X_n\downarrow X$ a.s.~and $X_n\leq c$ on $\{X<c\}$
% and
% $X_n\uparrow X$ a.s.~and $ X_n\geq c$ on $\{X\geq c\}$.
% Under this normalization, we then have
% \(
% f(X_n)\uparrow f(X)
% \) a.s.
% Thus, the monotone convergence theorem can be applied to extend the Jensen-gap representation to integrable random variables.
% We therefore introduce the following continuity condition for the extension on the integrable random variables.
% The bounded-continuity condition allows us to extend the representation to
% $\mathcal X_c(\tilde{\mathbb D})$. To further consider the larger domain
% $\mathcal X_1(\tilde{\mathbb D})$ of integrable random variables taking
% values in $\tilde{\mathbb D}$, bounded continuity is no longer sufficient,
% since the approximating random variables need not be uniformly bounded.
% We therefore introduce the following continuity condition.
% The idea is to approximate an integrable random variable $X$ radially from
% a constant $c\in\tilde{\mathbb D}$. More precisely, the
% approximating sequence remains between $c$ and $X$ and moves monotonically
% away from $c$ toward $X$. This motivates the following radial continuity
% condition (RC).
 
\begin{enumerate}
\item[(TC)] Truncation continuity: 
On the domain of $\rho$, if $a_n,b_n\in\mathbb D$,
$a_n\downarrow\inf\mathbb D$ and $b_n\uparrow\sup\mathbb D$, then $\rho(X_n)\to\rho(X)$ where $X_n=(X\vee a_n)\wedge b_n$.
\end{enumerate}

% \begin{enumerate}
% \item[(TC)] Truncation continuity: 
% On the domain of $\rho$,  $\rho(X_n)\rightarrow\rho(X)$ where $X_n=(X\vee n)\wedge(-n)$.
% \end{enumerate}

% When $\rho$ is interpreted as a measure of variability,
% the condition has a natural meaning: as the approximation gradually moves
% from the constant benchmark toward the full random variable $X$, the value
% of the functional is required to converge to $\rho(X)$. 
% The term ``radial'' indicates that $X_n$ remains between
% $c$ and $X$ and moves monotonically away from $c$ toward $X$. This condition is also consistent with a variability
% interpretation of $\rho$: when $\rho$ is viewed as a measure of
% variability, $X_n$ gradually moves from the constant benchmark $c$
% toward the random variable $X$, and the corresponding variability
% $\rho(X_n)$ is required to converge to $\rho(X)$.

We summarize the result in the next Proposition~\ref{thm:L1_BD}. We allow $\rho(X)$ to take the value
$ \infty$. This is because, although $X$ is integrable, $\E[f(X)]$ may be infinite
for a real-valued convex function $f$. Nevertheless, $\E[f(X)\mid\mG]$ remains well-defined in the extended sense. Since $f$ is convex on the open interval ${\mathbb D}$, for every $c\in{\mathbb D}$, there exists a subgradient $s$ such that $f(x)\geq f(c)+s(x-c)$ for every $x\in{\mathbb D}$. Hence, $f^-(x)=\max\{-f(x),0\}\leq\max\{kx+b,0\}$ for some $k,b\in \R$. We have $f^-(x)\leq|k||x|+| b|$, so $f^-(X)$ is integrable due to the integrability of $X$. Consequently, $\E[f(X)\mid\mG]$ is well-defined as an extended conditional expectation taking values in $(-\infty,\infty]$.
\begin{proposition}\label{thm:L1_BD}
    For $\rho:\mathcal X_1({\mathbb D})\rightarrow\R\cup\{\infty\}$, where ${\mathbb D}$ is an open interval and $\rho(X)\in\R$ for all $X\in\mathcal X_s({\mathbb D})$, the following statements are equivalent:
    \begin{enumerate}[label=(\roman*)]
\item $\rho$ satisfies (LI), (BD), (N), (CLS) and (TC);
\item $\rho$ is an $f$-Jensen gap on $\mathcal X_1(\mathbb D)$ for some convex function $f:\mathbb D\to\mathbb R$.
\end{enumerate} 
\end{proposition}
 The proof of Proposition~\ref{thm:L1_BD} is given in \ref{app:proof_chara_jens}.
 Via standard arguments, the condition (CLS) in Proposition~\ref{thm:L1_BD} can be replaced by
(CUS), noting that either condition is sufficient to obtain the Jensen-gap
representation on $\mathcal X_c(\mathbb D)$ by Proposition~\ref{thm:conti}. 
 
%  One may wonder why we do not simply impose the condition that
% \(
% X_n\downarrow X\) a.s.~on  $\{X<c\}$ and
% \(X_n\uparrow X\) a.s.~on $\{X\geq c\}$ without assuming the relation between $X_n$ and $c$.
% % imply $\rho(X_n)\to\rho(X)$. 
% Indeed, for each $\omega$ such that
% $X(\omega)>c$, if $X_n(\omega)\uparrow X(\omega)$, then there exists
% $N_\omega$ such that
% \(
% X_n(\omega)>c\) for \( n\geq N_\omega\).
% Hence, under the above normalization of $f$,
% $f(X_n(\omega))\uparrow f(X(\omega))$ for $n\geq N_\omega$.
% % A similar argument applies when $X(\omega)<c$. 
% The difficulty is that $N_\omega$ may depend on $\omega$, and there need
% not exist a deterministic $N$ such that
% \(
% f(X_n)\uparrow f(X)\) a.s.~for all \(n\geq N\).
% In fact, Example~\ref{ex:counter_L1_comp} provides a counterexample.
% Taking $c=1/2$, we have $X_n\uparrow X$ a.s., but
% \(
% \rho(X_n)\nrightarrow\rho(X).
% \) 

% Besides, one may wonder whether, by symmetry, the condition
% $|X_n-c|\uparrow|X-c|$ a.s.~can be replaced by
% $|X_n-c|\downarrow|X-c|$ a.s. However, this is generally not true,
% since the reverse monotone convergence theorem requires, for example,
% $f(X_1)\in L^1$. Again, Example~\ref{ex:counter_L1_comp} provides a
% counterexample. Taking $c=2$, we have
% $(X-c)(X_n-c)\geq0$ and $|X_n-c|\downarrow|X-c|$ a.s., but
% $\rho(X_n)\nrightarrow\rho(X)$.

The truncation continuity condition (TC) provides one direct route to extending the Jensen-gap representation to
$\mathcal X_1({\mathbb D})$. Besides, we give an alternative extension to
$\mathcal X_1({\mathbb D})$ based on the martingale continuity condition (MC) in \ref {app:L1_MC}. This approximation is constructed through conditional expectations with respect to increasingly informative sub-$\sigma$-algebras. 
\section{Identifying special forms of the generator}\label{s:identi_f}
Having characterized the general Jensen-gap representation in the previous section, we now investigate how additional properties of the functional $\rho$ restrict the form of its generator $f$.
 Here scaling invariance (SI) and translation invariance (TI) concern the ordering induced by $\rho$. In particular, (TI) is distinct from cash additivity and location invariance (LocI), which is introduced below. 
% At this stage, the generator is still largely unrestricted, apart from the fact that affine terms do not affect the Jensen gap, that is $\E[kX+b]-(k\E[X]+b)=0$. We therefore ask which natural consistency properties of $\rho$ can identify the nonlinear part of $f$.

The   starting point is consistency under scaling. When $\rho$ is interpreted as a measure of variability, a common change of scale should not reverse the ordering induced by $\rho$. In particular, if two random variables $X$ and $Y$ have the same mean and
$\rho(X)\geq\rho(Y)$, then under the transformations
$X\mapsto cX$ and $Y\mapsto cY$ for the same $c$, it is natural to require that the same ordering be preserved. The restriction to random variables with the same mean allows the comparison to focus on their dispersion rather than differences in their levels. This motivates the scaling invariance condition on $\rho$. As we will show, when \(\rho\) is a Jensen-gap functional with a measurable generator, the scaling invariance is strong enough to imply homogeneity of some degree \(\alpha\). Combining this homogeneity with the Jensen-gap representation identifies the logarithmic, \(x\log x\), and power forms of the generator up to an affine term.

Besides consistency under scaling, it is also natural to consider
consistency under translation. A common translation transforms $X$ and $Y$ into $X+c$ and $Y+c$ respectively. If $\rho$ is interpreted as a measure of variability, such
a common shift should not affect their relative ranking. Therefore, it is natural to require that their ordering under $\rho$ be preserved. We study this translation invariance condition after scaling invariance and show that it leads to a different class of generators. Now we first study scaling invariance and introduce the following two properties of $\rho$.
\begin{enumerate}
    \item[(SI)] Scaling invariance: for any $X,Y\in \mathcal X_s(\R_+)$ with $\E[X]=\E[Y]$, and any $c>0$, if
    $\rho(X)\geq\rho(Y)$, then
    \(
        \rho(cX)\geq\rho(cY).
    \)
    \item[(HDA)] Homogeneity of degree $\alpha$: $
\rho(cX)=c^\alpha\rho(X)
$
for all $c>0$ and all $X\in\mathcal X_s(\R_+)$.
\end{enumerate}
We first establish the following lemma to show the equivalence between (SI) and (HDA). 
\begin{lemma}\label{le:equi_si_hda}
    Assume that $\rho$ is an $f$-Jensen gap on $\mathcal X_s(\R_+)$ for some measurable $f:\mathbb \R_+\to\mathbb R$. Then the following
statements are equivalent:
\begin{enumerate}[label=(\roman*)]
    \item $\rho$ satisfies (SI);
    \item $\rho$ satisfies (HDA) for some $\alpha\in\R$.
\end{enumerate}
\end{lemma}

The proof of Lemma~\ref{le:equi_si_hda} is given in \ref{app:proof_identi_f}.
% The preceding lemma shows that the order-based scaling invariance condition (SI) is equivalent to homogeneity of some degree $\alpha$. 
% The preceding lemma shows that (SI) is equivalent to the existence of some
% $\alpha\in\R$ such that
% \(
% \rho(cX)=c^\alpha\rho(X), \qquad c>0.
% \)
We next use this homogeneity property (HDA) to further identify the generator
$f$. Suppose that
\(
\rho(X)=\E[f(X)]-f(\E[X]).
\)
Then (HDA) implies that
\(
\Psi_c(x)=f(cx)-c^\alpha f(x)
\) is affine. This leads to an identification of the nonlinear part of the generator $f$. More precisely, up to an
affine term, the possible forms of $f$ are
\begin{equation}\label{eq:form_f}
   f_{a,\alpha}(x)=\begin{cases}
       a\log x,\quad a\in\R\text{ for } \alpha=0,\\
       ax\log x,\quad a\in\R \text{ for } \alpha=1,\\
       ax^\alpha,\quad a\in\R\text{ for } \alpha\notin\{0,1\}.
   \end{cases} 
\end{equation}
This result is summarized in the following lemma.
\begin{lemma}\label{le:HDA_generator}
Fix $\alpha\in\R$. Assume that $\rho$ is an $f$-Jensen gap on $\mathcal X_s(\R_+)$ for some measurable $f:\mathbb \R_+\to\mathbb R$.
% Assume that there exists a measurable function
% $f: \R_+\to\R$ such that
% \(
% \rho(X)=\E[f(X)]-f(\E[X])
% \)
% for every $X\in\mathcal X_s(\R_+)$. 
Then the following
statements are equivalent:
\begin{enumerate}[label=(\roman*)]
    \item $\rho$ satisfies (HDA) with $\alpha$;
    \item there exist $a,k,b\in\R$ such that $f(x)=f_{a,\alpha}(x)+kx+b$ for all $x>0$. Equivalently, up to an affine term, the generator $f$ is of the form $f_{a,\alpha}$.
\end{enumerate}
\end{lemma}
The proof of Lemma~\ref{le:HDA_generator} is given in \ref{app:proof_identi_f}. The two preceding lemmas together yield the following characterization, which directly links scaling invariance to the possible forms of the generator $f$.
% Thus, (SI) ultimately determines the possible forms of
% the generator $f$. The result is stated in the following theorem.
\begin{theorem}\label{thm:rho_SI_form_f}
    Assume that $\rho$ is an $f$-Jensen gap on $\mathcal X_s(\R_+)$ for some measurable $f:\mathbb \R_+\to\mathbb R$. Then the following
statements are equivalent:
\begin{enumerate}[label=(\roman*)]
    \item $\rho$ satisfies (SI);
    \item up to an affine term, there exist some $a, \alpha\in\R$ such that $f=f_{a,\alpha}$.
\end{enumerate}
%  Equivalently, up to an affine term, the generator $f$ belongs to the
% class $f_{a,\alpha}$.
\end{theorem}
\begin{remark}\label{re:conn_varibility_measure}
    These results also connect our scaling invariance condition to measures of
variability. Here, we use the definition of the variability measure in
\citet[Definition~1]{bellini2022parametric}: a nonnegative law-invariant
functional that equals $0$ on constants and is homogeneous of some degree
$\alpha\geq0$. In our framework, law invariance (LI)
and binary decomposition (BD) first yield the Jensen-gap representation. The scaling invariance condition (SI)
then implies homogeneity of some degree $\alpha$. Therefore, together with
nonnegativity on $\rho$, whenever $\alpha\geq0$, the resulting Jensen-gap functional
is a variability measure.
\end{remark}
\begin{remark}\label{re:large_domain_f}
The positivity restriction in Lemma~\ref{le:equi_si_hda} can be removed. For example, suppose that $f:\R\to\R$ is measurable and that (SI) is correspondingly formulated for $X,Y\in\mathcal X_s(\R)$, with scaling factors $c\in\R$. Then this lemma remains valid in the sense that there exists some $\alpha\in\R$ such that
\(
\rho(cX)=|c|^\alpha\rho(X)
\)
for every $c\in\R\setminus\{0\}$ and $X\in\mathcal X_s(\R)$. We exclude $c=0$ because \(\alpha\) is unrestricted and \(|0|^\alpha\) may be undefined at the stage of Lemma~\ref{le:equi_si_hda}. The proof is analogous and is therefore omitted.  
\end{remark}
\begin{remark}
\label{re:large_domain_f2}
The choice of $(0,\infty)$ in Lemma~\ref{le:HDA_generator} provides a
common domain for $f$ under all values of $\alpha$. For a specific generator, the
domain and the admissible scaling factors $c$ may be enlarged accordingly.
For example, when $\alpha=2$ or $\alpha=3$, one may take the domain to be
$\R$ and formulate (HDA) for all $c\in\R$. In particular, for $\alpha=2$, if $f:\R\to\R$ is measurable and
\(
\rho(cX)=c^2\rho(X)
\)
for every $c\in\R$ and every $X\in\mathcal X_s(\R)$, then
the same argument as in Lemma~\ref{le:HDA_generator} yields that there exist $a,k,b\in\R$ such that
\(
f(x)=ax^2+kx+b
\)
for all $x\in\R$.
Similarly, for $f(x)=\sqrt{x}$, corresponding to $\alpha=1/2$, the simple random variables are restricted to take values in $[0,\infty)$ and the scaling factor satisfies $c\geq0$, so that $cX$ also takes values in $[0,\infty)$. Under these restrictions, the corresponding conclusion of Lemma~\ref{le:HDA_generator} remains valid.
\end{remark}
% \textcolor{red}{start writting here, formulate the whole story for translation invariance and write some collaries to identify variance, log, for variance it should have 2 corollary 1 from HDA, 1 from TI. What's more, rewrite the introduction of section 3.2}
Besides consistency under scaling, another natural requirement is consistency under translation. While scaling changes the unit of measurement, translation changes only the common level of the random variables. When $\rho$ is interpreted as a measure of variability, shifting two random variables by the same constant should not reverse the ordering induced by $\rho$. This motivates the following translation invariance condition, which is formulated in parallel with (SI). As in the scaling case, this order-based condition restricts the possible forms of the generator $f$. We therefore introduce the following two properties of $\rho.$
% which are related to the translation invariance condition.
\begin{enumerate}
    \item[(TI)] Translation invariance:
    for any $X,Y\in\mathcal X_s(\R)$ with
    $\E[X]=\E[Y]$, and any $c\in\R$, if
    \(
    \rho(X)\geq\rho(Y),
    \)
    then
    \(
    \rho(X+c)\geq\rho(Y+c).
    \)
    \item[(ET)] Exponential translation:
    there exists some $\beta\in\R$ such that
    \(
    \rho(X+c)=e^{\beta c}\rho(X)
    \)
    for all $c\in\R$ and all $X\in\mathcal X_s(\R)$.
\end{enumerate}
The first lemma connects the order-based condition (TI) with (ET).
\begin{lemma}\label{le:TI_ET}
Assume that $\rho$ is an $f$-Jensen gap on $\mathcal X_s(\R)$
for some measurable $f:\R\to\R$. Then the following statements are equivalent:
\begin{enumerate}[label=(\roman*)]
    \item $\rho$ satisfies (TI);
    \item $\rho$ satisfies (ET).
\end{enumerate}
\end{lemma}
% The proof of Lemma~\ref{le:TI_ET} is given in Appendix~\ref{app:proof_TI_ET}.
We next use the exponential translation property (ET) to further identify the generator $f$.
% Suppose that
% \(
% \rho(X)=\E[f(X)]-f(\E[X]).
% \)
% Then (ET) implies that, for every $c\in\R$,
% \[
% \Psi_c(x)=f(x+c)-e^{\beta c}f(x)
% \]
% is affine. This leads to an identification of the nonlinear part of the generator $f$.
% More precisely, up to an affine term,
Define
\begin{equation}\label{eq:TI_f}
    g_{a,\beta}(x)
=
\begin{cases}
ae^{\beta x}, & a\in\R,\quad \beta\neq0,\\[1mm]
ax^2, & a\in\R,\quad \beta=0.
\end{cases}
\end{equation}
The possible forms of $f$ are summarized in the following lemma.

\begin{lemma}\label{le:ET_form_f}
Fix $\beta\in\R$. Assume that $\rho$ is an $f$-Jensen gap on $\mathcal X_s(\R)$
for some measurable $f:\R\to\R$. Then the following statements are equivalent:
\begin{enumerate}[label=(\roman*)]
    \item $\rho$ satisfies (ET) with $\beta$;
    \item there exist $a,k,b\in\R$ such that
    \(
    f(x)=g_{a,\beta}(x)+kx+b
    \)
    for all $x\in\R$. Equivalently, up to an affine term, the generator $f$ is of the form $g_{a,\beta}$.
\end{enumerate}
\end{lemma}
The two preceding lemmas together yield the following characterization, which directly links translation invariance to the possible forms of the generator $f$.
\begin{theorem}\label{thm:TI_form_f}
Assume that $\rho$ is an $f$-Jensen gap on $\mathcal X_s(\R)$
for some measurable $f:\R\to\R$. Then the following statements are equivalent:
\begin{enumerate}[label=(\roman*)]
    \item $\rho$ satisfies (TI);
    \item up to an affine term, there exist some $a, \beta\in\R$ such that $f=g_{a,\beta}$.
    % there exist $\beta,a,k,b\in\R$ such that
    % \[
    % f(x)=g_{a,\beta}(x)+kx+b
    % \]
    % for all $x\in\R$.
\end{enumerate}
\end{theorem}
The proofs of Lemma~\ref{le:TI_ET}, Lemma~\ref{le:ET_form_f}, and Theorem~\ref{thm:TI_form_f} are analogous to those of Lemma~\ref{le:equi_si_hda}, Lemma~\ref{le:HDA_generator}, and Theorem~\ref{thm:rho_SI_form_f}, respectively, and are therefore omitted.
The preceding results allow us to give direct characterizations of two
important Jensen-gap functionals: the logarithmic Jensen gap and the
variance. In each case, the general Jensen-gap structure is first determined
by law invariance and the decomposition principle, while an additional
invariance property identifies the specific generator. The nonnegativity (N) here is to guarantee that $f$ is convex on the corresponding interval and thus measurable.
\begin{theorem}\label{thm:log_characterization}
For $\rho:\mathcal X_s(\R_+)\to\R$, the following statements are equivalent:
\begin{enumerate}[label=(\roman*)]
    \item $\rho$ satisfies (LI), (BD), (N), and
    \(
    \rho(cX)=\rho(X)
    \)
    for every $c>0$ and every $X\in\mathcal X_s(\R_+)$;

    \item $\rho$ is an $f$-Jensen gap on $\mathcal X_s(\R_+)$ for some
$f=a\log$ with $a\leq0$.
\end{enumerate}
\end{theorem}
 
\begin{remark}\label{re:camp_wong_log}
\citet{campbell2025excessgrowth} provide three axiomatic characterizations of the excess growth rate, which is the Jensen-gap functional with a logarithmic generator. Their Theorem~3.1 uses a chain rule together with homogeneity of degree $0$ and other conditions to characterize the logarithmic Jensen gap, while their Theorem~3.11 obtains the general Jensen-gap structure under an affine condition (D4), which prescribes affine dependence on the probabilities when the mean is fixed. In our approach, we first characterize the general Jensen-gap functionals through law invariance and the binary decomposition principle, without prescribing this dependence. Our scaling invariance condition then implies homogeneity of some degree $\alpha$, which identifies a broader class of generators, including the logarithmic, $x\log x$, and power generators. %A detailed comparison is given in \ref{app:campbell_wong}.
\end{remark}

After the logarithmic case, we turn to another fundamental Jensen-gap
functional, the variance. Variance is both location invariant (exponential translation with $\beta=0$) and homogeneous of degree two. Moreover,
Theorems~\ref{thm:rho_SI_form_f} and~\ref{thm:TI_form_f} show that, among
Jensen-gap functionals, variance is the only one, up to a multiplicative
constant, that satisfies both scaling invariance and translation invariance.
Thus, requiring the ordering induced by $\rho$ to be preserved under affine
transformations $X\mapsto kX+b$ identifies the quadratic generators. In addition, from Proposition~\ref{prop:JG-convex} and its proof, either quasi-convexity or convexity of $\rho$ provides another characterization of variance. We now
introduce the relevant properties and summarize the resulting characterization
of variance in Theorem~\ref{thm:variance_characterization}.  
% The following theorem
% shows that, within the Jensen-gap class characterized above, either property
% is sufficient to identify the variance.
% We first introduce the following two properties for the variance:
\begin{enumerate}
    \item[(LocI)] Location invariance:
    \(
    \rho(X+c)=\rho(X)
    \)
    for every $c\in\R$ and every $X\in\mathcal X_s(\R)$;

    \item[(HDA$_2$)] Homogeneity of degree two:
    \(
    \rho(cX)=c^2\rho(X)
    \)
    for every $c\in\R$ and every $X\in\mathcal X_s(\R)$.
    \item[(ATI)] Affine transform invariance: for any $X,Y\in\mathcal X_s(\R)$ with $\E[X]=\E[Y]$, and for any $k,b\in \R$, if $\rho(X)\geq\rho(Y)$, then $\rho(kX+b)\geq\rho(kY+b)$.
%     \item[(QCx)]  Quasi-convexity:
%  $
%     \rho(tX+(1-t)Y)\le\max\{\rho(X),\rho(Y)\}
%  $   for every $X,Y\in\mathcal X_s(\R)$
% and every $t\in[0,1]$.
%     \item[(Cx)] Convexity:
% \(
% \rho(tX+(1-t)Y)
% \leq t\rho(X)+(1-t)\rho(Y)
% \)
\end{enumerate}
% \textcolor{red}{variance is the only one both satisfies SI and TI, so its order is preserved under transform kX+b,formulate it here and add it to the following thm}
\begin{theorem}\label{thm:variance_characterization}
For $\rho:\mathcal X_s(\R)\to\R$, the following statements are equivalent:
\begin{enumerate}[label=(\roman*)]
    \item $\rho$ satisfies (LI), (BD), (N) and (LocI);
\item $\rho$ satisfies (LI), (BD), (N) and (HDA$_2$);
\item $\rho$ satisfies (LI), (BD), (N) and (ATI);
\item $\rho$ satisfies (LI), (BD), (N) and  is quasi-convex;
\item $\rho$ satisfies (LI), (BD), (N) and is convex;
    \item $\rho$ is an $f$-Jensen gap on $\mathcal X_s(\R)$ for some $f(x)=ax^2$ with $a\geq0$, that is, $\rho=a\var$.
\end{enumerate}
\end{theorem}
The proof of Theorem~\ref{thm:variance_characterization} is given in \ref{app:proof_identi_f}. Under smoothness assumptions on the generator \(f\), \citet[Theorem~4.4]{Chafai2006} implies a characterization of variance through the convexity of Jensen-gap functionals, which has been discussed in our Remark~\ref{re:chafai_cx}. Our result establishes this characterization without requiring smoothness of \(f\) as we only require it to be convex via (N). Moreover, the equality \(\rho(X+c)=\rho(X)\) in (LocI) can be replaced by the condition \(\rho(X+c)\leq\rho(X)\). Indeed, replacing \(X\) by \(X+c\) and \(c\) by \(-c\) yields \(\rho(X)\leq\rho(X+c)\) and hence equality follows.

\begin{remark}\label{re:var_log_large_domain}
    The characterization theorems above
(Theorems~\ref{thm:log_characterization} and
\ref{thm:variance_characterization}) can also be extended to larger
domains of random variables. For example, by imposing semicontinuity, the domain can be enlarged to compactly supported random variables, which is similar to Proposition~\ref{thm:conti}. Moreover, by imposing truncation continuity (TC), the characterization can be further extended to
integrable random variables, as in Proposition~\ref{thm:L1_BD}.

% For example, to obtain the corresponding characterization on integrable
% random variables, we may proceed in the same way as in
% Proposition~\ref{thm:L1_BD}. Assume that $\rho(X)<\infty$ for corresponding simple random variables $X$. The characterization result can be applied on this subclass to obtain
% the Jensen-gap representation. The additional invariance or homogeneity condition then determines the specific form of the generator $f$. Finally, the corresponding continuity condition extends
% the identified Jensen-gap representation from simple random variables to
% the larger domain of integrable random variables.
\end{remark}

After seeing the form
$$
\rho(X)= \E[f(X)]-f(\E[X]),
$$
one may expect the decomposition principle
 $\rho(X)=\E[\rho(X\mid\mathcal G)]+\rho(\E[X\mid\mathcal G])$ 
 to hold for a general sub-$\sigma$-algebra $\mathcal G$ which is not finite. 
 % In particular, this automatically holds true if we define $\rho(X\mid\mathcal G)=\E[f(X)\mid\mathcal G]-f(\E[X\mid\mG])$ for general $\mathcal G$. 
 However, such extensions of the above characterization theorems require a formal definition of the conditional functional $\rho(X\mid\mathcal G)$ for any sub-$\sigma$-algebra $\G$. We formalize this in the next section.
\section{Extensions to general sigma-algebras}\label{s:cond_fun_general_G}
% Let $(\Omega,\mathcal F,\mathbb{P})$ be an atomless probability space. In this section, we give the definition of the conditional version of functions on the distribution of a random variable. And based on this definition, the conditional versions of some common functions are rigorously proved, such as $\mathbb E[f(X)]$ and $\text{ES}(X)$.
% In this section, we extend the definition of conditional law-invariant
% functionals to an arbitrary sub-$\sigma$-algebra
% $\mG\subseteq\mathcal F$ and establish the measurability of the resulting
% conditional functionals. We then verify that this construction recovers
% the usual conditional forms of several familiar functionals, including
% $\mathbb E[f(X)]$ and $\operatorname{ES}(X)$. In addition, we discuss which characterization results in Section~\ref{s:jensen_func} continue to hold when the corresponding decomposition principle is formulated with respect to more general sub-$\sigma$-algebras.
In this section, we extend the framework developed for finite
sub-$\sigma$-algebras to general sub-$\sigma$-algebras. In Subsection~\ref{ss:con_genral_G}, we formulate
conditional law-invariant functionals in this general setting and illustrate
the construction with two familiar examples, $\E[f(X)]$ and the expected
shortfall (ES).  We also give a dynamic interpretation of conditional Jensen gaps. In Subsection~\ref{ss:gener_sigma_chara}, we investigate
which characterization results in Section~\ref{s:jensen_func} remain valid
under decomposition principles formulated with respect to general
sub-$\sigma$-algebras.
\subsection{Conditional risk measures with a general sigma-algebra}\label{ss:con_genral_G}
In this subsection, we recall the evaluation $\sigma$-algebra on the space of probability measures \citep{georgii2011gibbs} and we use it to formulate conditional versions of law-invariant functionals with respect to general sub-$\sigma$-algebras. We also verify the measurability of the resulting conditional functionals, and illustrate the construction through the familiar examples $\E[f(X)]$ and Expected Shortfall $\ES_\alpha(X)$.
\begin{definition}[Evaluation $\sigma$-algebra on probability measures] 
Let $S$ be a Borel subset of $\mathbb R$, and let $\mathcal M(S)$ denote the set of all probability measures on $\mathcal B(S)$, where $\mathcal B(S)$ denote the Borel $\sigma$-algebra on $S$. For each $B\in\mathcal B(S)$, define the evaluation map $$e_B:\mathcal M(S)\to[0,1], \qquad e_B(\mu)=\mu(B).$$ Then the evaluation $\sigma$-algebra on $\mathcal M(S)$ is defined by 
\begin{equation}
    \begin{aligned}
       \Sigma_{\mathcal M} &= \sigma\bigl(e_B: B\in\mathcal B(S)\bigr) = \sigma\bigl\{e^{-1}_B(\tilde{B}):B\in\mathcal B(S),\,\tilde{B}\in\mathcal B\left([0,1]\right)\bigr\}\\&=\sigma\left\{
\{\mu:\mu\in\mathcal M(S)\text{ and }\mu(B)\in\tilde B\}:
B\in\mathcal B(S),\,\tilde B\in\mathcal B([0,1])
\right\}.
    \end{aligned}
\end{equation}
Equivalently, $\Sigma_{\mathcal M}$ is the smallest $\sigma$-algebra on $\mathcal M(S)$ such that, for every Borel set $B\in\mathcal B(S)$, $e_B$ is measurable. 
% In particular, if $\nu$ is a map from a measurable space $(\Omega,\mathcal H)$ into $\mathcal M_1(S)$, then \[ \nu:(\Omega,\mathcal H)\to(\mathcal M_1(S),\Sigma_{\mathcal M}) \] is measurable if and only if, for every $B\in\mathcal B(S)$, the map \[ \omega\mapsto \nu(\omega)(B) \] is $\mathcal H$-measurable. 
\end{definition}
% \begin{definition}[Law invariant functional]
% Let $\Gamma$ be a functional defined on a class of $S$-valued random variables.
% We say that $\Gamma$ is law invariant if
% \[
% X\overset{d}=Y
% \quad\Longrightarrow\quad
% \Gamma(X)=\Gamma(Y).
% \]
% Equivalently, there exists a map
% \[
% \widetilde{\Gamma}:\mathcal M(S)\to\mathbb R
% \]
% such that
% \[
% \Gamma(X)=\widetilde{\Gamma}(\mu_X),
% \]
% where $\mu_X=\mathbb P\circ X^{-1}$ denotes the distribution of $X$.
% \end{definition}
% \begin{definition}[Measurability of $\widetilde{\Gamma}$]
% We say that the distributional version $\widetilde{\Gamma}$ is measurable if
% \[
% \widetilde{\Gamma}:
% (\mathcal M_1(S),\Sigma_{\mathcal M})
% \to
% (\mathbb R,\mathcal B(\mathbb R))
% \]
% is measurable.
% \end{definition}
\begin{definition}[Conditional version of a law-invariant functional $\nu$]\label{def:con_fun}
Let $\mathcal G\subseteq\mathcal F$ be a sub-$\sigma$-algebra, and let
$\mu_{X\mid\mathcal G}$ be a conditional distribution of $X$ given
$\mathcal G$, that is,
\[
\mu_{X\mid\mathcal G}(\omega,B)
=
\p(X\in B\mid\mathcal G)(\omega)=\E[\mathbf 1_{\{X\in B\}}\mid \mG](\omega)\quad\text{a.s.},
\qquad B\in\mathcal B(S).
\]
Besides, $\mu_{X\mid\mathcal G}(\cdot,B)$ is $\mathcal G$-measurable for every
$B\in\mathcal B(S)$, and $\mu_{X\mid\mathcal G}(\omega,\cdot)$ is a probability measure on $\mathcal B(S)$ for every $\omega\in\Omega$.
Let \(\mathcal P\) be a measurable subset of \(\mathcal M(S)\), equipped with the trace sigma-algebra \(\Sigma_{\mathcal M}|_{\mathcal P}=\{A\cap\mathcal P:A\in\Sigma_{\mathcal M}\}\). Assume that
$
\tilde{\nu}:(\mathcal P,\Sigma_{\mathcal M}|_{\mathcal P})
\to \left(\RR,\mathcal B(\RR)\right)
 $ 
is measurable and \(\nu(X)=\tilde{\nu}(\mu_X)\). Whenever \(\mu_{X\mid\mathcal G}(\omega,\cdot)\in\mathcal P\) a.s., define
\[
\nu(X\mid\mathcal G)(\omega)
=\tilde{\nu}\bigl(\mu_{X\mid\mathcal G}(\omega,\cdot)\bigr).
\]
% Assume that \(
% \tilde{\nu}:
% (\mathcal M(S),\Sigma_{\mathcal M})
% \to
% (\mathbb R,\mathcal B(\mathbb R))
% \) 
% is measurable and $\nu(X)=\tilde{\nu}(\mu_{X})$. We define the conditional
% version of $\nu$ given $\mathcal G$ by
% \[
% \nu(X\mid\mathcal G)(\omega)
% =
% \tilde{\nu}\bigl(\mu_{X|\mathcal G}(\omega,\cdot)\bigr).
% \]
\end{definition} 
This is well-defined, since whenever $\mu_{X\mid\mathcal G}$
takes values in $\mathcal P$ a.s., we can choose a version
that does so everywhere.
% \begin{definition}[Conditional version of $\Gamma$]\label{def:con_fun}
% Let $X$ be an $S$-valued random variable and let $\mathcal G\subseteq\mathcal F$ be a sub-$\sigma$-algebra. Let
% $\mu_{X|\mathcal G}$ be a regular conditional distribution of $X$ given $\mathcal G$. That is, $\mu_{X|\mathcal G}$ is a probability kernel from $(\Omega,\mathcal G)$ to $(S,\mathcal B(S))$ such that, for every $B\in\mathcal B(S)$, the map
% $\omega\mapsto \mu_{X|\mathcal G}(\omega,B)$
% is $\mathcal G$-measurable and is a version of
% $\mathbb E[\mathbf 1_{\{X\in B\}}\mid\mathcal G]$. Equivalently, for every $B\in\mathcal B(S)$,
% $$
% \mu_{X|\mathcal G}(\omega,B)
% =
% \mathbb P(X\in B\mid\mathcal G)(\omega)
% =
% \mathbb E[\mathbf 1_{\{X\in B\}}\mid\mathcal G](\omega)
% \quad\text{a.s.}
% $$
% Moreover, for every $\omega\in\Omega$, the map
% $B\mapsto \mu_{X|\mathcal G}(\omega,B)$
% is a probability measure on $(S,\mathcal B(S))$.

% Assume that
% $$
% \widetilde{\Gamma}:(\mathcal M(S),\Sigma_{\mathcal M})
% \to
% (\mathbb R,\mathcal B(\mathbb R))
% $$
% is measurable. We define the conditional version of $\Gamma$ given $\mathcal G$ by
% $$
% \Gamma(X\mid\mathcal G)(\omega)
% :=
% \widetilde{\Gamma}\bigl(\mu_{X|\mathcal G}(\omega,\cdot)\bigr).
% $$
% \end{definition}
\begin{proposition}\label{thm1:measurable_con_func}
   The conditional version of $\nu$ defined above is measurable with respect to $\mathcal G $.
\end{proposition}
The proof of Proposition~\ref{thm1:measurable_con_func} is given in \ref{app:proof_gener_G}.
Given the definition of the conditional version of general functionals, we want to show that some common law-based functionals are measurable under our framework.
% \begin{lemma}\label{le:mean_meas}
%     Let $f: S\to\mathbb R$ be Borel measurable and define
% $\Sigma_{\mathcal M|D_f}=\{\mu\in\mathcal M(S):\int |f|d\mu<\infty\}$. Then $\Sigma_{\mathcal M|D_f}$ is a $\sigma$-algebra and $\tilde{\Gamma}(\mu_X)=\mathbb E[f(X)]$ is measurable to $\Sigma_{\mathcal M|D_f}$.
% \end{lemma}
% \begin{proof}
%     We first show that $\Sigma_{\mathcal M|D_f}$ is a $\sigma$-algebra
% \end{proof}
% \begin{lemma}\label{le:mean_meas}
% Let $f:S\to\mathbb R$ be Borel measurable and define
% $D_f:=\{\mu\in\mathcal M(S):\int f^+\,\text{d}\mu<\infty\text{ or }\int f^-\,\text{d}{\mu}<\infty\}$, where $f^+=\max(f,0)$ and $f^-=\max(-f,0)$.
% Then $D_f\in\Sigma_{\mathcal M}$, and the functional
% $\tilde{\Gamma}_f:D_f\to\mathbb R$ defined by
% $\tilde{\Gamma}_f(\mu):=\int f\,\text{d}\mu$
% is measurable with respect to the trace $\sigma$-algebra
% $\Sigma_{\mathcal M}|_{D_f}:=\{A\cap D_f:A\in\Sigma_{\mathcal M}\}$.
% In particular, if $X$ is an $S$-valued random variable with law $\mu_X\in D_f$, then
% $\tilde{\Gamma}_f(\mu_X)=\mathbb E[f(X)]$.
% \end{lemma}
The following lemma is a standard consequence of the measurability of kernels \citep[Lemma~3.1]{kallenberg2021}.
\begin{lemma}\label{le:mean_meas}
Let $f:S\to\mathbb R$ be Borel measurable and define
$D_f=\{\mu\in\mathcal M(S):\int |f|\,\d\mu<\infty\}$.
Then $D_f\in\Sigma_{\mathcal M}$, and the functional
$\tilde{\nu}_f:D_f\to\mathbb R$ defined by
$\tilde{\nu}_f(\mu)=\int f\,\d\mu$
is measurable with respect to the trace $\sigma$-algebra
$\Sigma_{\mathcal M}|_{D_f}=\{A\cap D_f:A\in\Sigma_{\mathcal M}\}$.
% In particular, if $X$ is an $S$-valued random variable with law $\mu_X\in D_f$, then
% $\tilde{\nu}_f(\mu_X)=\mathbb E[f(X)]$.
\end{lemma}
The proof of Lemma~\ref{le:mean_meas} is given in \ref{app:proof_gener_G}.
The following proposition shows that, for integral-type functionals, the law-based conditional version recovers the usual conditional expectation. This is a standard consequence of the disintegration formula for conditional distributions \citep[Theorem~8.5]{kallenberg2021}. 
% We only show the sketch of the proof.
%We include the proof for completeness.
\begin{proposition}\label{prop:con_mean}
With the notation of Lemma~\ref{le:mean_meas}, let $X$ be an $S$-valued random variable with law $\mu_X$, and let $\mathcal G\subseteq\mathcal F$ be a sub-$\sigma$-algebra. Assume that $\mu_X\in D_f$.
Then
$\mu_{X\mid\mathcal G}(\omega,\cdot)\in D_f$ a.s.~and
$\tilde{\nu}_f(\mu_{X\mid\mathcal G})
=\int f\,\d\mu_{X\mid\mathcal G}
=\E[f(X)\mid\mathcal G]$ a.s.
\end{proposition}
The sketch of the proof of Proposition~\ref{prop:con_mean} is given in \ref{app:proof_gener_G}.
\begin{remark}\label{re:extend_R}
    The domain $D_f$ in Lemma~\ref{le:mean_meas} and Proposition~\ref{prop:con_mean} can be relaxed to
    $$
    D_f^{\mathrm{ext}}
    =
    \left\{
    \mu\in\mathcal M(S):
    \int f^+\,\d\mu<\infty
    \text{ or }
    \int f^-\,\d\mu<\infty
    \right\},
    $$
    provided that \(\tilde{\nu}_f\) is regarded as a functional \(\tilde{\nu}_f:D_f^{\mathrm{ext}}\to\RR\).
    For $\mu\in D_f^{\mathrm{ext}}$, define
    $
    \tilde{\nu}_f(\mu)
    =
    \int f\,\d\mu
    =
    \int f^+\,\d\mu
    -
    \int f^-\,\d\mu,
    $ which is well-defined since the case \(\infty-\infty\) is excluded.
    With this convention, the conclusions of Lemma~\ref{le:mean_meas} and Proposition~\ref{prop:con_mean} remain valid in the extended sense.
    % $$
    % \tilde{\Gamma}_f(\mu_{X|\mathcal G})
    % =
    % \mathbb E[f(X)\mid\mathcal G] \text{ a.s.},
    % $$
    % where both sides may take values in $\overline{\mathbb R}$.
\end{remark}
The previous proposition shows that the law-based conditional version of
an integral functional recovers the usual conditional expectation. As an
illustration, we consider the Expected Shortfall. Recall that, for
$\alpha\in(0,1)$ and $X\in L^1$,
\begin{equation}\label{eq:es_def}
    \operatorname{ES}_\alpha(X)
    =
    \min_{m\in\mathbb R}
    \left\{
        m+\frac{1}{1-\alpha}\E[(X-m)^+]
    \right\}.
\end{equation}
% see \citet{rockafellar_uryasev_2002}.
\begin{example}\label{prop:con_ES}
Let $\alpha\in(0,1)$, $X$ be integrable and $\mG$ be any sub-$\sigma$-algebra. Let
$K_\omega=\mu_{X\mid\mathcal G}(\omega,\cdot)$. Then the law-based
conditional Expected Shortfall  has the standard representation \citep[Example~6.2]{RuszczynskiShapiro2006}
\[
    \operatorname{ES}_\alpha(X\mid\mathcal G)
    =
    q_\alpha
    +
    \frac{1}{1-\alpha}
    \E[(X-q_\alpha)^+\mid\mathcal G]
    =
    \operatorname*{ess\,inf}_{Z\in L^1(\mathcal G)}
    \left\{
        Z+\frac{1}{1-\alpha}
        \E[(X-Z)^+\mid\mathcal G]
    \right\}
    \quad\text{a.s.},
\]
where
\begin{equation}\label{eq:q_alpha}
        q_\alpha(\omega)
    =
    \inf\{t\in\mathbb R:K_\omega((-\infty,t])\geq\alpha\}
    \in L^1(\mathcal G).
\end{equation}
A self-contained proof of the equalities in Example~\ref{prop:con_ES} is given in \ref{app:proof_gener_G}.
\end{example}
\begin{remark}\label{re:bounded_ES}
If $X\in L^\infty(\mathcal F)$, then $q_\alpha\in L^\infty(\mathcal G)$.
Hence,
\[
    \operatorname{ES}_\alpha(X\mid\mathcal G)
    =
    \operatorname*{ess\,inf}_{Z\in L^\infty(\mathcal G)}
    \left\{
        Z+\frac{1}{1-\alpha}
        \E[(X-Z)^+\mid\mathcal G]
    \right\}
    \quad\text{a.s.}
\]
\end{remark}

The conditional version of the Jensen gap also has a dynamic interpretation through the decomposition principle, which we illustrate next.
\begin{proposition}\label{cor:dynamic_jensen}
Let $\mathbb D$ be an open interval, $f:\mathbb D\to\R$ be convex, and $(\mathcal F_t)_{t\in[0,T]}$ be a filtration with $\mathcal F_t\subseteq\mathcal F$. For $X\in\mathcal X_1(\mathbb D)$ with $f(X)\in L^1$, define
\[
D_t(X)=\E[f(X)\mid\mathcal F_t]-f(\E[X\mid\mathcal F_t]).
\]
Then $D_t(X)$ is nonnegative and integrable, and, for $0\leq s\leq t\leq T$,
\[
D_s(X)=\E[D_t(X)\mid\mathcal F_s]+D_s(\E[X\mid\mathcal F_t])\quad\text{a.s.}
\]
In particular, $(D_t(X))_{t\in[0,T]}$ is a nonnegative supermartingale.
\end{proposition}
The proof is given in \ref{app:proof_gener_G}. Here $D_t(X)$ measures the remaining variability of $X$ given the information at time $t$. Its conditional expected value cannot increase as information arrives.

\subsection{Extensions of the characterization theorems}\label{ss:gener_sigma_chara}
In this subsection, we discuss how the characterization results in
Section~\ref{s:jensen_func} can be extended when the decomposition
principle is imposed with respect to general sub-$\sigma$-algebras. Assume that there exists a measurable functional
$\tilde\rho:(\mathcal P,\Sigma_{\mathcal M}|_{\mathcal P})
\to\R\cup\{\infty\}$ such that
$\rho(X)=\tilde\rho(\mu_X)$, where
$\Sigma_{\mathcal M}|_{\mathcal P}
=\{A\cap\mathcal P:A\in\Sigma_{\mathcal M}\}$ and $\mu_X$ is the law of $X$.
Let
\[
\mathcal P_s=\{\mu_X:X\in\mathcal X_s(\mathbb D)\},\qquad
\mathcal P_c=\{\mu_X:X\in\mathcal X_c(\mathbb D)\},\qquad
\mathcal P_1=\{\mu_X:X\in\mathcal X_1(\mathbb D)\}.
\]
Here, $\mathcal P_s$ is the law domain corresponding to the simple random
variables $\mathcal X_s(\mathbb D)$ considered in
Theorems~\ref{thm:rho_form_simple}, \ref{thm:rho_form_nonne_convex},
\ref{thm:log_characterization}, and \ref{thm:variance_characterization}.
Similarly, $\mathcal P_c$ is the law domain corresponding to the $\mathbb D$-compactly supported
random variables $\mathcal X_c(\mathbb D)$ considered in
Proposition~\ref{thm:conti}, while $\mathcal P_1$ is the law domain corresponding to the
integrable random variables $\mathcal X_1(\mathbb D)$ considered in
Proposition~\ref{thm:L1_BD}. 

For
\(
\rho(X)
=
\E[\rho(X\mid\mG)]
+
\rho(\E[X\mid\mG])
\)
to be well-defined, two issues need to be considered. Firstly,
$\rho(X\mid\mG)$ should be $\mG$-measurable. Secondly, the conditional law of $X$ given $\mG$ and the law of the conditional expectation
$\E[X\mid\mG]$ should remain in the corresponding law domain $\mathcal P$. Note that the conditional law and conditional expectation are defined
only up to a.s.~equality. Hence, if they belong to their respective
domains a.s., we may choose versions that do so everywhere.
% Let the law domains $\mathcal P_s$ correspond to the simple random variables $\mathcal X_s(\mathbb D)$ in Theorems~\ref{thm:rho_form_simple}, \ref{thm:rho_form_nonne_convex}, \ref{thm:log_characterization} and \ref{thm:variance_characterization},
% $\mathcal P_c$ correspond to the $\mathbb D$-compactly supported random variables $\mathcal X_c({\mathbb D})$ in Proposition~\ref{thm:conti}, and $\mathcal P_1$ correspond to the integrable random variables $\mathcal X_1({\mathbb D})$ in Proposition~\ref{thm:L1_BD}. 
The measurability requirement is ensured by the following lemma.

\begin{lemma}\label{le:law_domain_meas}
The law domains $\mathcal P_s$, $\mathcal P_c$, and $\mathcal P_1$
are measurable subsets of their corresponding spaces of
probability measures, that is, they are in their corresponding evaluation $\sigma$-algebra $\Sigma_{\mathcal M}$. Consequently, if
$\mu_{X\mid\mG}$ is in the relevant law
domain $\mathcal P$, then
\(
\rho(X\mid\mG)
=
\tilde\rho(\mu_{X\mid\mG})
\)
is $\mG$-measurable.
\end{lemma}
The proof of Lemma~\ref{le:law_domain_meas} is given in \ref{app:proof_gener_G}.
However, the measurability of the law domains alone is not sufficient for the decomposition principle to be well-defined. We also need to ensure that conditioning preserves the relevant domain. More precisely, for
\(
\rho(X)
=
\E[\rho(X\mid\mG)]
+
\rho(\E[X\mid\mG])
\)
to be meaningful, the conditional law of $X$ and the law of $\E[X\mid\mG]$ should remain in the law domain $\mathcal P$.
% while $\E[X\mid\mG]$ should admit a version belonging to the domain of $\rho$.

For Theorems~\ref{thm:rho_form_simple}, \ref{thm:rho_form_nonne_convex}, \ref{thm:log_characterization} and \ref{thm:variance_characterization}, condition (BD) cannot in general
be extended from binary $\sigma$-algebras to arbitrary sub-$\sigma$-algebras. The corresponding law domain is $\mathcal P_s$. Indeed, if $X\in\mathcal X_s(\mathbb D)$ takes finitely many values
$x_1,\ldots,x_n$ lying in the interval $\mathbb D$, then conditioning only changes the
probabilities assigned to these values. Hence, the conditional law of
$X$ given $\mG$ remains in the law domain $\mathcal P_s$. However,
$\E[X\mid\mG]$ need not be simple. For example, let $U$ and $V$ be independent uniformly distributed random
variables on $(0,1)$ and define
\(
    X=\id_{\{V\leq U\}}\) where \(
    \mG=\sigma(U).
\)
Then $X$ is a simple random variable with
\(
    \p(X=1)=\E[U]=\frac12,
\)
whereas
\(
    \p(X=1\mid\mG)=U\) and \(
    \E[X\mid\mG]=U
\) a.s.
Thus, the law of $\E[X\mid\G]$ is not in $\mathcal P_s$. 
% Hence, $\rho(\E[X\mid\mG])$ need not be well-defined when the
% domain of $\rho$ is $\mathcal X_s(\mathbb D)$.

In contrast, for Propositions~\ref{thm:conti} and \ref{thm:L1_BD}, the condition (BD) can be generalized as follows.
\[
    \rho(X)
    =
    \E[\rho(X\mid\mG)]
    +
    \rho\bigl(\E[X\mid\mG]\bigr)\quad\text{for every sub-$\sigma$-algebra $\mG\subseteq\mathcal F$}
    \tag{GD}
\]
For
Proposition~\ref{thm:conti}, the law domain is $\mathcal P_c$. If
\(
a\leq X\leq b
\)
a.s.~for some $[a,b]\subseteq{\mathbb D}$, then
\(
\mu_{X\mid\mG}([a,b])=\E[\id_{\{a\leq X\leq b\}}\mid\mG]=1\) a.s.
and
\(
a\leq\E[X\mid\mG]\leq b
\)
a.s. Thus, the conditional law of
$X$ given $\mG$ and the law of $\E[X\mid\G]$ in (GD) both remain in $\mathcal P_c$.

For Proposition~\ref{thm:L1_BD}, the law domain is $\mathcal P_1$, which corresponds to the laws of the integrable random variables taking values in ${\mathbb D}$. If $X\in\mathcal X_1({\mathbb D})$, then
Proposition~\ref{prop:con_mean} gives
\(
\int |x|\,\mu_{X\mid\mG}(\d x)<\infty\) a.s. That is, the conditional law has a finite first moment a.s. Moreover,
$\E[X\mid\mG]$ is integrable and takes values in
${\mathbb D}$ a.s. Hence, the conditional law of
$X$ given $\mG$ and the law of $\E[X\mid\G]$ in (GD) both remain in $\mathcal P_1$.
\section{Conclusion}\label{s:conclu}
Jensen gaps provide a useful framework that covers many important quantities,
including variance, excess growth rate, several measures of
variability, Bregman information, $f$-divergences and so on. Some existing characterization results focus on Jensen gaps for
two-valued simple random variables \citep{ebanks2008jensen,olbrys2021taffine,olbrys2022affine}.
More recently, \citet{campbell2025excessgrowth} study broader
characterizations, but their results either concern a particular generator
or impose conditions that already closely reflect the Jensen-gap form.

In this paper, we first study the basic properties of Jensen gaps, focusing on the connections between convexity-related properties and the generator \(f\). Then we introduce conditional versions of law-invariant functionals with respect to finite sub-$\sigma$-algebras, which are essential for formulating the binary decomposition principle. Next, we show that binary
decomposition (BD) and law invariance (LI) characterize the Jensen-gap representation on
simple random variables. Under suitable additional conditions, the
representation extends to larger classes of random variables, including compactly supported ones and integrable ones. We further show that scaling invariance (SI) and
translation invariance (TI) provide two alternative ways to identify the
generator: (SI) leads to logarithmic, $x\log x$, and power generators,
whereas (TI) leads to exponential and quadratic generators. This yields direct characterizations of the logarithmic Jensen gap and
the variance. We
also extend the conditional framework and the corresponding results from
finite sub-$\sigma$-algebras to general ones.
\section*{Acknowledgements}
Ruodu Wang is supported by the Natural Sciences and Engineering Research Council of Canada (CRC-2022-00141, RGPIN-2024-03728). GPT 5--6 were used in preparing this manuscript. The authors are responsible for the final content.
\appendix
\renewcommand{\thesection}{Appendix \Alph{section}}

\setcounter{theorem}{0}
\setcounter{lemma}{0}
\setcounter{proposition}{0} 
\setcounter{corollary}{0}
\renewcommand{\thelemma}{S.\arabic{lemma}}
\renewcommand{\thetheorem}{S.\arabic{theorem}}
\renewcommand{\theproposition}{S.\arabic{proposition}}
\renewcommand{\thecorollary}{S.\arabic{corollary}}
\setcounter{equation}{0}
\renewcommand{\theequation}{S.\arabic{equation}}
\providecommand{\theHequation}{}
\renewcommand{\theHequation}{S.\arabic{equation}}
% \section{Further extensions of the Jensen-gap characterization}\label{app:extensions}
% This appendix collects further extensions of the Jensen-gap characterization developed in the main text. In \ref{app:L1_MC}, we provide an alternative route to the integrable domain
% $\mathcal X_1(\tilde{\mathbb D})$ based on the martingale continuity condition (MC). 
% Unlike the radial approximation used in the main text, this approach approximates a random variable by conditional expectations with respect to an increasing sequence of finite sub-$\sigma$-algebras, thereby exploiting progressively richer information. 
% In \ref{app:2n_equi}, We develop a separate characterization on the more restrictive class
% $\mathcal X_e(\mathbb D)$ of $2^n$-equiprobable simple random variables. In this setting, the binary decomposition principle (BD) is replaced by the weaker equiprobable binary decomposition principle (EBD), and we show that this restricted structure still leads to the Jensen-gap representation. Corresponding extensions of the equiprobable characterization to broader domains are also provided.
\section{Extension by martingale continuity}\label{app:L1_MC}
In the main text, we extend the Jensen-gap representation to
$\mathcal X_1(\mathbb D)$ through the truncation continuity (TC).
Here, we provide an alternative extension based on a different type of approximation. We approximate $X$ by its conditional expectations with respect to an increasing sequence of finite sub-$\sigma$-algebras
$\{\mathcal G_n\}$. As these $\sigma$-algebras reveal progressively more information about $X$, the conditional expectations
$\E[X\mid\mathcal G_n]$ provide increasingly accurate approximations of $X$.
This motivates the following martingale continuity condition.
\begin{enumerate}
\item[(MC)] Martingale continuity:
On the domain of $\rho$, if $\{\mathcal G_n\}$ is a sequence of finite sub-$\sigma$-algebras satisfying
$\mathcal G_n\subseteq\mathcal G_{n+1}$ and $X$ is measurable with respect to
$\sigma\left(\bigcup_{n=1}^{\infty}\mathcal G_n\right)$, then
\(
\rho\left(\E[X\mid\mathcal G_n]\right)\to\rho(X).
\)
\end{enumerate}
The condition (MC) also admits a natural interpretation in terms of information and variability. When $\rho$ is interpreted as a measure of variability, the conditional expectations
$\E[X\mid\mathcal G_n]$ capture the variability of $X$ under increasingly rich information. Thus, as the available information becomes sufficient to recover $X$, the corresponding values
$\rho(\E[X\mid\mathcal G_n])$ are required to converge to the variability
$\rho(X)$. We now present an alternative result that extends the Jensen-gap characterization to
$\mathcal X_1({\mathbb D})$ under the martingale continuity condition (MC) in the following.
\begin{proposition}\label{thm:L1_MC}
    For $\rho:\mathcal X_1({\mathbb D})\rightarrow\R\cup\{\infty\}$, where ${\mathbb D}$ is an open interval and $\rho(X)\in\R$ for all $X\in\mathcal X_s({\mathbb D})$, the following statements are equivalent:
    \begin{enumerate}[label=(\roman*)]
\item $\rho$ satisfies (LI), (BD), (N) and (MC);
\item $\rho$ is an $f$-Jensen gap on $\mathcal X_1(\mathbb D)$ for some convex function $f:\mathbb D\to\mathbb R$.
\end{enumerate} 
\end{proposition}
\begin{proof}
    (ii)$\Rightarrow$(i): (LI) and (N) hold trivially. (BD) follows from the proof of Proposition~\ref{thm:L1_BD}. Now it remains to show (MC). Similar to the proof of Proposition~\ref{thm:L1_BD}, we can construct a convex function $g$ such that $g$ is nonnegative, convex and $\rho(X)=\E[f(X)]-f(\E[X])=\E[g(X)]-g(\E[X])$. Let $\E[X\mid\G_n]=M_n$. From the conditional Jensen's inequality, we have $g(M_n)\leq\E[g(X)\mid\G_n]$ a.s. Thus $\E[g(M_n)]\leq\E[g(X)]$. And by L\'evy's upward theorem, we have $M_n=\E[X\mid\G_n]\to\E\left[X\mid\sigma\left(\bigcup_{n=1}^\infty \G_n\right)\right]=X$ a.s. Together with the continuity of $g$, we have $g(M_n)\to g(X)$ a.s. By Fatou's lemma, we have $\E[g(X)]\leq\liminf_{n\to\infty}\E[g(M_n)]$. Therefore, we have $\lim_{n\to\infty}\E[g(M_n)]=\E[g(X)]$. Since $\E[M_n]=\E[X]$, we obtain that 
    \begin{equation*}
        \begin{aligned}
            \rho(M_n)&=\E[f(M_n)]-f(\E[M_n])=\E[g(M_n)]-g(\E[M_n])\\&\rightarrow \E[g(X)]-g(\E[X])=\E[f(X)]-f(\E[X]).
        \end{aligned}
    \end{equation*}
     (i)$\Rightarrow$(ii): From (LI), (BD), (N) and $\rho(X)\in\R$ for $X\in\mathcal X_s({\mathbb D})$, we can have that $\rho(X)=\E[f(X)]- f(\E[X])$ holds for every $X\in\mathcal X_s({\mathbb D})$ from Theorem~\ref{thm:rho_form_nonne_convex}, and $f$ is convex on ${\mathbb D}$. Now we use (MC) to extend the result to $\mathcal X_1({\mathbb D})$. Let $\mathbb Q=\{q_1,q_2,\ldots\}$ denote the rational numbers, then $\sigma(X)=\sigma\left(\{X<q_i\}:i\geq1\right)$. Define $\G_n=\sigma\left(\{X<q_i\}:1\leq i\leq n\right)$. Then $\sigma(\bigcup_{n=1}^\infty \G_n)=\sigma(X)$. For every $X\in\mathcal X_1({\mathbb D})$, we have $\rho\left(\E[X\mid\G_n]\right)\to\rho(X)$ by (MC). Let $M_n=\E[X\mid\G_n]$. Since $\G_n$ is finite, $M_n\in\mathcal X_s({\mathbb D})$. Thus, $\rho\left(\E[X\mid\G_n]\right)=\rho(M_n)=\E[f(M_n)]- f(\E[M_n])$. Similar to the proof of (ii)$\Rightarrow$(i), we can have $\rho(M_n)=\E[f(M_n)]- f(\E[M_n])\to\E[f(X)]- f(\E[X])$, thus we obtain that $\rho(X)=\E[f(X)]- f(\E[X])$ for every $X\in\mathcal X_1({\mathbb D})$.
\end{proof}
\section{Proofs of  results in Section~\ref{s:jensen_gap}}\label{app:proof-2}

% We first prove Proposition~\ref{prop:JG-convex} from Section~\ref{s:jensen_gap}, which establishes the connections between various convexity-related properties of Jensen-gap functionals and the convexity of their generator \(f\).
\begin{proof}[Proof of Proposition~\ref{prop:JG-convex}.]
Convexity of $f$ directly gives  {(ii)--(viii)},
by Jensen's inequality and the definition of $J_f$.
Conversely, fix $x,y\in\mathbb R$ and $\lambda\in[0,1]$,
and let $Z$ take $x,y$ with probabilities $\lambda,1-\lambda$.
Each of (ii)--(iv) implies
\[
0\le J_f(Z)
=\lambda f(x)+(1-\lambda)f(y)
-f(\lambda x+(1-\lambda)y),
\]
and thus convexity  of $f$ in (i). 
To see this, for (iii), apply mixture concavity to point masses;
for (iv), use $\E[Z]\le_{\mathrm{cx}}Z$.
Thus (i)--(iv) are equivalent. 
Next,  each of (v)--(vii) trivially implies (viii).
It then suffices
 to prove (viii)$\Rightarrow$(i). For this, fix the mean $c$ in (viii). 
For arbitrary $x,y\in\mathbb R$, put
$m=(x+y)/2$ and $z=2c-m$.
Choose $(X,Y)$ taking the values
$(x,y)$, $(y,x)$, and $(z,z)$ with probabilities
$1/4$, $1/4$, and $1/2$, respectively.
Then $\E[X]=\E[Y]=c$ and $X\laweq Y$.
Quasi-convexity gives
\[
J_f\left(\frac{X+Y}{2}\right)
\le \max\{J_f(X),J_f(Y)\}=J_f(X).
\] 
Hence
\[
0\le J_f(X)-J_f\left(\frac{X+Y}{2}\right)
=\frac14\bigl(f(x)+f(y)-2f(m)\bigr).
\]
All quantities are finite because these variables are simple.
Thus $f$ is midpoint convex, and its measurability implies
convexity. This proves the  equivalences between {(i)--(viii)}.

For the last part, note that the implications  (ix)$\Rightarrow$(x)$\Rightarrow$(xi) are immediate. 
It suffices to show (xi)$\Rightarrow$(ix). 
By the first part, $f$ is convex and hence continuous. 
For any simple random variable $X$, $c\in\R$, and $t\in(0,1)$, $X+c$ is a convex combination of $X/t$ and $c/(1-t)$. Hence,
quasi-convexity, nonnegativity, and $J_f(c)=0$ for constants $c$ together give
\[
J_f(X+c)
\le \max\left\{J_f(X/t),J_f\left(\frac{c}{1-t}\right)\right\}
=J_f(X/t).
\]
Since $X$ is simple and $f$ is continuous, letting $t\uparrow1$ yields
$J_f(X+c)\le J_f(X)$.
Applying this inequality to $X+c$ and $-c$ gives the reverse inequality.
Therefore $J_f(X+c)=J_f(X)$ for every simple $X$ and every $c$.

Fix $c\in\mathbb R$ and define $h_c(x)=f(x+c)-f(x)$.
For every two-point random variable $X$,
 $
J_{h_c}(X)=J_f(X+c)-J_f(X)=0.
 $ 
Taking arbitrary two-point distributions gives
\[
h_c(tx+(1-t)y)=t h_c(x)+(1-t)h_c(y),
\qquad x,y\in\mathbb R,\quad t\in[0,1].
\]
Hence $h_c$ is affine, so
 $
f(x+c)=f(x)+f(c)-f(0)+k(c)x
 $
for some function $k$. Interchanging $x$ and $c$ gives
$k(c)x=k(x)c$. Taking $x=1$, we obtain $k(c)=k (1) c$.
Write $a=k(1)/2$, and we now have 
$$
f(x+c)=f(x)+f(c)-f(0)+2a c x,\qquad \mbox{$c,x\in \R$}.
$$  
Set $v(x)=f(x)-f(0)-ax^2$. The identity becomes
$v(x+c)=v(x)+v(c)$, so $v(r)=rv(1)$ for every rational $r$.
By continuity, this holds for every real $r$. Hence
 $
f(x)=ax^2+v(x) +f(0)
 $ with $v$ linear. Convexity of $f$ gives $a\ge0$. Thus 
 $J_f=a\var$ on $L^q$.  
\end{proof}

\section{Proofs of  results in Section~\ref{s:jensen_func}}\label{app:proof_chara_jens}
\begin{proof}[Proof of Theorem~\ref{thm:rho_form_simple}]
(ii)$\Rightarrow$(i): (LI) is trivial. For (BD): Since $\rho$ is a law invariant functional on $\mathcal X_s(\mathbb D)$, from Definition~\ref{def:con_risk_meas_fini}, we obtain that
    \begin{align*}
\E\bigl[\rho(X\mid\mathcal{G})
\bigr]+\rho\bigr(\E[X\mid\mathcal{G}]\bigl)
&=
\E\Bigl[
\E\bigl[f(X)\mid\mathcal{G}\bigr]
-f\bigl(\E[X\mid\mathcal{G}]\bigr)
\Bigr] +\E [f (\E[X\mid\mathcal{G}] ) ]
-f (\E [\E[X\mid\mathcal{G}] ] ) \\
&=
\E\bigl[f(X)\bigr]-f\bigl(\E[X]\bigr) = 
\rho(X).
\end{align*}

(i)$\Rightarrow$(ii): 
Let $x,y\in \mathbb D$ and $s,r,t\in[0,1]$.
Consider a simple random variable
\[
Z=
\begin{cases}
x, & \text{with probability } sr+(1-s)t,\\
y, & \text{with probability } 1-sr-(1-s)t.
\end{cases}
\]
Let $U_1$ and $U_2$ be two independent uniformly distributed random variables on $[0,1]$. Define
$
A_1=\{U_1\le s\},
$
$
A_2=\{U_2\le r\}
$
,
$
A_3=\{U_2\le t\}
$
,
$
X_1
=
\id_{A_2}x
+
\id_{A_2^c}y
$ and $
X_2
=
\id_{A_3}x
+
\id_{A_3^c}y.
$
Since $X_1$ and $X_2$ are independent of $A_1$, we can have
$
Z
\laweq
\id_{A_1}X_1
+
\id_{A_1^c}X_2.
$
Hence by (LI), $\rho(Z)=\rho(\id_{A_1}X_1
+
\id_{A_1^c}X_2)$. Let $\mG=\sigma(A_1)$.
From Definition~\ref{def:con_risk_meas_fini}, we have $\rho(\id_{A_1}X_1
+
\id_{A_1^c}X_2\mid \mG)=\id_{A_1}\rho(X_1)+\id_{A_1^c} \rho(X_2)$ a.s.  
Define \(D(x,y;t)=\rho(\tilde X)\), where \(\tilde X\) has the distribution \(t\delta_x+(1-t)\delta_y\). 
%where $\tilde X$ is a two-valued random variable satisfying
%$
%P(\tilde X=x)=t$, and $
%P(\tilde X=y)=1-t.
%$
Then, by (BD),
\begin{align*}
\rho(Z)
&=
D\bigl(x,y;sr+(1-s)t\bigr)=\rho(\id_{A_1}X_1
+
\id_{A_1^c}X_2)
\\
&=
s\rho(X_1)
+
(1-s)\rho(X_2)
+
\rho\Bigl(
\id_{A_1}\E[X_1]
+
\id_{A_1^c}\E[X_2]
\Bigr)
\\
&=
sD(x,y;r)
+
(1-s)D(x,y;t)
+
D\bigl(
rx+(1-r)y,\,
tx+(1-t)y;\,
s
\bigr).
\end{align*}
   Rearranging the terms yields
\begin{equation}\label{eq:key_eq_jen}
\begin{aligned}
&D\bigl(
rx+(1-r)y,\,
tx+(1-t)y;\,
s
\bigr)
\\ &=
D\bigl(
x,y;\,
sr+(1-s)t
\bigr) 
-sD(x,y;r)
-(1-s)D(x,y;t),
\end{aligned}
\end{equation}
for all $x,y\in \mathbb D$, $r,t,s\in[0,1]$.  
% and $s\in(0,1)$. Note when $s=0$ or $s=1$, Eq.~\eqref{eq:key_eq_jen} also holds, since both LHS and RHS equal to 0. Hence, Eq.~\eqref{eq:key_eq_jen} holds for all $x,y\in \mathbb D$ and $s,r,t\in[0,1]$.
Therefore, by \citet[Theorem~1]{olbrys2022affine},
there exists a function $f:\mathbb D\to\mathbb R$ such that
$
D(x,y;t)
=
tf(x)
+(1-t)f(y)
-f\bigl(tx+(1-t)y\bigr)$
for every $x,y\in\mathbb D$ and $t\in[0,1]$. This means that
\(
\rho(X)=\E[f(X)]-f(\E[X])
\) for every two-valued simple random variable $X$ taking values in $\mathbb D$.
%     Now for n-valued simple random variable, we prove it by induction. Suppose for a simple random variable taking at most $n$ values Eq.~\eqref{eq:rho_form1} holds, then we show for simple random variable taking $n+1$, it still holds.\\
%     Let $X^*$ take values $\{x_1,x_2,...,x_{n+1}\}$, $P(X^*=x_i)=t_i$ and $\sum_{i=1}^{n+1}t_i=1$. W.L.O.G, we assume $t_i\in(0,1) \text{ for }\forall i\in\{1,2,...,n+1\}$, otherwise, $X^*$ reduces to a simple random variable taking $n$ values. 
%     Define $A=\{U_1> t_{n+1}\}$, $B_i=\{\frac{\sum_{k=1}^{i-1}t_k}{1-t_{n+1}}< U_2\leq\frac{\sum_{k=1}^{i}t_k}{1-t_{n+1}}\}$, where we let $t_0=0$. Then $P(A)=1-t_{n+1}$ and $P(B_i)=\frac{t_i}{t_{n+1}}$ for $i\in\{1,2,...,n\}$. Let $Y^*=\sum_{i=1}^n \mathbf{1}_{B_i} x_i$. Due to the independency of $U_1$ and $U_2$, we have $Y^*$ is independent of $A$. Then we can have
%     $$X^*\overset{d}=\mathbf{1}_{A}Y+\mathbf{1}_{A^c}x_{n+1}.$$
%     Taking $\mathcal G=\sigma(A)$, then we obtain
%     \begin{equation*}
%         \begin{aligned}
% \rho(X^*)&=\rho(\mathbf{1}_{A}Y+\mathbf{1}_{A^c}x_{n+1})\overset{(iii)}=E[\mathbf{1}_A\rho(Y)+\mathbf{1}_{A^c}\rho(x_{n+1})]+\rho(\mathbf{1}_AE(Y)+\mathbf{1}_{A^c}x_{n+1})\\&\overset{\text{by induction}}=(1-t_{n+1})\left(E[f(Y)]-f(E[Y])\right)+(1-t_{n+1})f(E[Y])\\&+t_{n+1}f(x_{n+1})-f\left((1-t_{n+1})E[Y]+t_{n+1}x_{n+1}\right)\\&=\sum_{i=1}^{n+1}t_if(x_i)-f(\sum_{i=1}^{n+1}t_ix_i)=E[f(X^*)]-f(E[X^*]).
%         \end{aligned}
%     \end{equation*}
% \end{proof}

Now let $X^*$ be a simple random variable taking at most $n+1$ values. We prove the result by induction. Assume that $\rho$ is an $f$-Jensen gap for every simple random variable taking at most $n$ values. Let $X^*$ take values $\{x_1,x_2,\ldots,x_{n+1}\}$ with
$
\p(X^*=x_i)=t_i$, $ i=1,\ldots,n+1,
$
where $\sum_{i=1}^{n+1} t_i=1$. Without loss of generality, assume that
$
t_i\in(0,1)$, $ i=1,\ldots,n+1$, since otherwise $X^*$ reduces to a simple random variable taking at most $n$ values. Define
$
A=\{U_1>t_{n+1}\},
$
and
$
B_i=
\left\{
\sum_{k=0}^{i-1}t_k/(1-t_{n+1})
<
U_2
\le
\sum_{k=0}^{i}t_k/(1-t_{n+1)}
\right\},
\,i=1,\ldots,n,
$
where $t_0=0$. 
% Then
% \[
% \mathbb P(A)=1-t_{n+1},
% \qquad
% \mathbb P(B_i)=\frac{t_i}{1-t_{n+1}},
% \quad i=1,\ldots,n.
% \]
Let
$
Y^*=\sum_{i=1}^{n}\mathbf 1_{B_i}x_i.
$
% Since $U_1$ and $U_2$ are independent, $Y^*$ is independent of $A$. Moreover,
% \[
% \mathbb P(Y^*=x_i)=\frac{t_i}{1-t_{n+1}},
% \qquad i=1,\ldots,n.
% \]
% Hence,
We have
\(
X^*
\laweq
\mathbf 1_A Y^*
+
\mathbf 1_{A^c}x_{n+1}.
\)
Taking $\mathcal G=\sigma(A)$ and applying Condition (BD), we obtain that
\[
\begin{aligned}
\rho(X^*)
&=
\rho\!\left(
\mathbf 1_A Y^*
+
\mathbf 1_{A^c}x_{n+1}
\right) \\
&=
\E\!\left[
\mathbf 1_A\rho(Y^*)
+
\mathbf 1_{A^c}\rho(x_{n+1})
\right]
+
\rho\!\left(
\mathbf 1_A\E[Y^*]
+
\mathbf 1_{A^c}x_{n+1}
\right) \\
&=
(1-t_{n+1})
\bigl(
\E[f(Y^*)]-f(\E[Y^*])
\bigr)
+
(1-t_{n+1})f(\E[Y^*]) \\
&\qquad
+
t_{n+1}f(x_{n+1})
-
f\left(
(1-t_{n+1})\E[Y^*]
+
t_{n+1}x_{n+1}
\right)\\&=\sum_{i=1}^{n+1} t_i f(x_i)
-
f\!\left(
\sum_{i=1}^{n+1} t_i x_i
\right) \\
&=
\E[f(X^*)]-f(\E[X^*]).
\end{aligned}
\] This completes the induction.
\end{proof}
% \subsection{Proof of Proposition~\ref{thm:conti}}\label{app:proof_rho_bdd}
Before we prove Proposition~\ref{thm:conti}, we introduce the following lemma.
\begin{lemma}\label{le:binary_finite}
    For $\rho:\mathcal X_c({\mathbb D})\rightarrow\R$, suppose $\rho$ satisfies the binary decomposition (BD) and law invariance (LI), then it satisfies the decomposition principle $\rho(X)=\E[\rho(X\mid\G)]+\rho(\E[X\mid\G])$ for any finite sub-$\sigma$-algebra $\G$.
\end{lemma}
\begin{proof}[Proof of Lemma~\ref{le:binary_finite}]
Let $A_1,\ldots,A_n$ be the non-null atoms of $\mathcal G$.
Let $p_i=\p(A_i)>0$ and $m_i=\E[X\mid A_i]$ for $i=1,\ldots,n$. Define $X_0=X$ and $X_i=m_i\id_{A_i}+X_{i-1}\id_{A_i^c}$ for $i=1,...,n$. Define $\mathcal H_i=\sigma(A_i)$. By binary decomposition, we have that $\rho(X_{i-1})=\rho(\E[X_{i-1}\mid \mathcal H_i])+p_i\rho(X_{i-1}\mid A_i)+(1-p_i)\rho(X_{i-1}\mid A_i^c)$. Note that $\E[X_i\mid\mathcal H_i]=\E[X_{i-1}\mid\mathcal H_i]$ and $\rho(m)=0$ for constant $m$, thus by (BD) we obtain that $\rho(X_{i})=\rho(\E[X_{i-1}\mid \mathcal H_i])+(1-p_i)\rho(X_{i-1}\mid A_i^c)$. Note that on the event $A_i$, we have that $X_{i-1}=X$. Hence, we obtain that $\rho(X_{i-1})=\rho(X_i)+p_i\rho(X_{i-1}\mid A_i)=\rho(X_i)+p_i\rho(X\mid A_i)$. Recursively, we obtain that $\rho(X)=\rho(X_0)=\rho(X_{n})+\sum_{i=1}^{n}p_i\rho(X\mid A_i)$. Note $X_{n}=\E[X\mid \G]$ a.s., thus we obtain that $\rho(X)=\rho(\E[X\mid\G])+\E[\rho(X\mid\G)]$.
\end{proof}
\begin{proof}[Proof of Proposition~\ref{thm:conti}] 
    (iii)$\Rightarrow${(i)--(ii)}: (LI), (BD) and (N) hold trivially. As for (CLS) and (CUS), we know that a convex function is continuous on an open interval, thus the bounded convergence theorem can be applied to obtain that $\lim_{n\rightarrow\infty}\rho(X_n)=\rho(X)$. 
    
    (i)$\Rightarrow$(iii): By Theorem~\ref{thm:rho_form_nonne_convex}, there exists some convex function \(f:\mathbb D\to\mathbb R\) such that $\rho(X)=\E[f(X)]-f(\E[X])$ holds for every $X\in\mathcal X_s({\mathbb D})$. Now for any $X\in \mathcal X_c({\mathbb D})$, there exists $[a_X,b_X]\subseteq{\mathbb D}$ such that $a_X\le X\le b_X$ a.s. Enumerate the rational numbers as $\mathbb Q=\{q_1,q_2,...\}$. Then $\sigma(X)=\sigma\left(\{X<q_i\}:i\geq1\right)$. Define $\G_n=\sigma\left(\{X<q_i\}:1\leq i\leq n\right)$. Let $X_n=\E[X\mid\G_n]$. Then $X_n\in\mathcal X_s({\mathbb D})$, $a_X\leq X_n\leq b_X$ a.s.~and $\rho(X_n)=\E[f(X_n)]-f(\E[X_n])$. Note that $\sigma(\bigcup_{n=1}^\infty \G_n)=\sigma(X)$. Hence, by L\'evy's upward theorem, we have that $X_n\rightarrow X$ a.s. By Lemma~\ref{le:binary_finite}, we obtain that $\rho(X)=\rho(\E[X\mid\G_n])+\E[\rho(X\mid\G_n)]$. By (N), we have 
    \begin{equation}\label{eq:key_rho1}
        \rho(X)\geq\rho(\E[X\mid\G_n])=\rho(X_n)
    \end{equation}
    
    Suppose (CLS) holds, so we have that $\rho(X)\leq\liminf_{n\rightarrow\infty}\rho(X_n)$. From $\rho(X)\geq\rho(X_n)$, we obtain that $\rho(X)\geq\limsup_{n\rightarrow\infty}\rho(X_n)$. Therefore, we obtain that $\rho(X)=\lim_{n\rightarrow\infty}\rho(X_n)=
\lim_{n\rightarrow\infty} \E[f(X_n)]-f(\E[X_n])=\E[f(X)]-f(\E[X])$.

 (ii)$\Rightarrow$(iii):
Suppose (CUS) holds. Define $J_f(X)=\E[f(X)]-f(\E[X])$ and $R(X)=\rho(X)-J_f(X)$. Since $\rho(X)\geq\rho(X_n)$ from \eqref{eq:key_rho1}, we obtain that $\rho(X)\geq\lim_{n\rightarrow\infty}\rho(X_n)=J_f(X)$. Thus $R(X)\geq0$. It remains to show that $R(X)=0$. Suppose not, then $R(X)>0$. Let $m_1=(a_X+b_X)/2$, $A=\{a_X\leq X\leq m_1\}$ and $p=\p(A)$. $\E[X\mid\sigma(A)]$ is a simple random variable, so $\rho\left(\E[X\mid\sigma(A)]\right)=J_f\left(\E[X\mid\sigma(A)]\right)$. Therefore, by (BD), we obtain that $R(X)=\E[\rho(X\mid\sigma(A))]-\E[J_f(X\mid\sigma(A))]=p R(X\mid A)+(1-p) R(X\mid A^c)$. 
% Hence, at least one of $R(X\mid A)$ and $R(X\mid A^c)$ is no less than $R(X)$. Without loss of generality, suppose that
% \(
% R(X\mid A)\ge R(X).
% \)
% Choose $Y_1$ such that
% \(
% \mu_{Y_1}=\mu_{X\mid A}.
% \)
% Then $Y_1\in I_1=[a_X,m_1]$ a.s. and
% \(
% R(Y_1)\ge R(X)>0.
% \)
Hence, there exists $B\in\{A,A^c\}$ with $\p(B)>0$ such that $R(X\mid B)\geq R(X)$. We consider the case $B=A$ and the case $B=A^c$ is analogous. Choose $Y_1$ such that $\mu_{Y_1}=\mu_{X\mid B}$ and $Y_1\in I_1=[a_X,m_1]$ a.s. Then we have $R(Y_1)\geq R(X)>0$. 
Similarly, we can split $Y_1$ at the midpoint $m_2=(a_X+m_1)/2$. Then we obtain $Y_2$ such that $R(Y_2)\geq R(Y_1)\geq R(X)>0$ and $Y_2\in I_2$ a.s., where $I_2=[a_X,m_2]$ or $[m_2,m_1]$. Repeating the similar arguments at each step, we construct a sequence $(Y_n)_{n\ge1}$ and a
sequence of nested closed intervals $(I_n)_{n\ge1}$ such that
$R(Y_n)\ge R(X)>0$, $Y_n\in I_n$ a.s.~for every $n$, and
the lengths of $I_n$ converge to $0$. Thus there exists $x\in[a_X,b_X]$ such that $|Y_n-x|\leq(b_X-a_X)/2^n$ a.s. Hence $Y_n\rightarrow x$
a.s. By the bounded convergence theorem, we obtain that $J_f(Y_n)\rightarrow J_f(x)=0$. By (CUS), we have that $\limsup_{n\rightarrow\infty}\rho(Y_n)\leq\rho(x)=0$. Therefore, $\limsup_{n\rightarrow\infty} R(Y_n)\leq0$, which contradicts $R(Y_n)\geq R(X)>0$.
\end{proof}
% \subsection{Proof of Proposition~\ref{thm:L1_BD}}\label{app:proof_rho_L1_RC}
\begin{proof}[Proof of Proposition~\ref{thm:L1_BD}]
(ii)$\Rightarrow$(i): (LI), (N) and (CLS) hold trivially. For (TC), since $\mathbb D$ is an open interval, there exists a positive integer $k$ such that $a_k<b_k$. Choose any $c\in(a_k,b_k)$. 
Since $f$ is convex on the open interval ${\mathbb D}$, there exists a subgradient $s$ such that $f(x)\geq f(c)+s(x-c)$ for every $x\in{\mathbb D}$. Define \begin{equation*}
    g(x)=f(x)-s(x-c)-f(c).
\end{equation*}
Then $g(x)$ is convex on ${\mathbb D}$, $g(x)\geq0$ and $g(x)$ attains its minimum 0 at $c\in{\mathbb D}$. Hence, $g(x)$ is decreasing in ${\mathbb D}\cap(-\infty,c)$ and increasing on ${\mathbb D}\cap[c,\infty)$. If $X\in[a_k,b_k]$, then $X_n=X$ for $n\geq k$. If $X<a_k$, then $X_n\downarrow X$ for $n\geq k$. If $X\geq b_k$, then $X_n\uparrow X$ for $n\geq k$. Therefore, $g(X_n)\uparrow g(X)$ for $n\geq k$. Thus, by the monotone convergence theorem, allowing the limit to be $\infty$, we have that $\E[g(X_n)]\rightarrow \E[g(X)]$. For $n\geq k$, $X_n$ lies between $c$ and $X$, so $|X_n|\leq |X|+2|c|$. $X$ is integrable and $X_n\rightarrow X$ a.s., so by the dominated convergence theorem, $\E[X_n]\rightarrow\E[X]$. Together with the continuity of $g$,
% Choose any $a_n$ and $b_n$ such that $l<a_n<c<b_n<r$, $a_n\downarrow l$, and $b_n\uparrow r$, we have $T_n(X(\omega))\downarrow X(\omega)$ if $X(\omega)< c$ and $T_n(X(\omega))\uparrow X(\omega)$ if $X(\omega)\geq c$. 
% For any $\{X_n\}$ such that $(X-c)(X_n-c)\geq 0$ a.s. and $|X_n-c|\uparrow|X-c|$ a.s., we have $X_n\downarrow X$ and $X_n\leq c$ on $X<c$, and $X_n\uparrow X $ and $X_n\geq c$ on ${X\geq c}$ a.s. And since $g$ is convex on the open interval ${\mathbb D}$, it is continuous, $g(X_n)\uparrow g(X)$ a.s. Thus, by the monotone convergence theorem, allowing the limit to be $+\infty$, we have that $\E[g(X_n)]\rightarrow \E[g(X)]$. As $|X_n-c|\leq |X-c|$, we have $|X_n|\leq |X-c|+|c|\leq |X|+2|c|$. Since $X$ is integrable, and $X_n\rightarrow X$ a.s., by dominated convergence theorem, we have $\E[X_n]\rightarrow\E[X]\in\R$. Hence, 
we have $g(\E[X_n])\rightarrow g(\E[X])\in \R$. 
Thus we obtain
\begin{align*}
    \rho(X_n)& =\E[f(X_n)]-f\left(\E[X_n]\right) \\ & =\E[g(X_n)]-g\left(\E[X_n]\right)\\&\rightarrow \E[g(X)]-g\left(\E[X]\right)=\E[f(X)]-f(\E[X])=\rho(X).
\end{align*}
Now for (BD), as here we allow $\infty$, we need to verify that (BD) holds under this extended case. From above, we know that $\rho(X)=\E[g(X)]-g(\E[X])$, where $g:{\mathbb D}\to[0,\infty)$ is convex. Let $Y=\E[X\mid \mG]$ and $H=\E[g(X)\mid \mG]-g(Y)\in[0,\infty]$. Hence, $H+g(Y)=\E[g(X)\mid \mG]$ and all three random variables in this identity are nonnegative. Therefore, taking expectations and using the additivity of the
Lebesgue integral for nonnegative measurable functions yields $\E[H]+\E[g(Y)]=\E[g(X)]$, where the equality is understood in $[0,\infty]$. Hence, $\E[H]+\E[g(Y)]-g(\E[X])=\E[g(X)]-g(\E[X])=\rho(X)$. Note $\E[\rho(X\mid \mG)]=\E[\E[g(X)\mid\mG]-g(\E[X\mid\mG])]=\E[H]$ and $\rho(\E[X\mid\mG])=\E[g(\E[X\mid\mG])]-g(\E[X])=\E[g(Y)]-g(\E[X])$. Therefore, under the extended case, we have shown that $\rho(X)=\rho(\E[X\mid\mG])+\E[\rho(X\mid\mG)]$ without the issue of $\infty-\infty$.

(i)$\Rightarrow$(ii): From (LI), (BD), (N), (CLS), and $\rho(X)\in\R$ for
$X\in\mathcal X_s(\mathbb D)$, by arguments similar to those in
Proposition~\ref{thm:conti}, we obtain that there exists a convex function
$f:\mathbb D\to\R$ such that
$\rho(X)=\E[f(X)]-f(\E[X])$ for every
$X\in\mathcal X_c(\mathbb D)$. Now we use condition (TC) to extend the result to any $X\in\mathcal X_1({\mathbb D})$. Choose any sequences $(a_n)$ and $(b_n)$ satisfying (TC). Then we have $\rho(X_n)\rightarrow\rho(X)$. Since $X_n\in\mathcal X_c(\mathbb D)$, by an argument similar to
(ii)$\Rightarrow$(i), we obtain that $\rho(X_n)=\E[f(X_n)]-f(\E[X_n])\rightarrow\E[f(X)]-f(\E[X])$. Therefore, $\rho(X)=\E[f(X)]-f(\E[X])$ for all $X\in\mathcal X_1(\mathbb D)$.
% that is, we need to construct a sequence of simple random variables $\{X_n\}$ satisfying (RC). Define $\tau_n(u)=\left(2^{-n}\lfloor2^n u\rfloor\right)\wedge n$. Then $\tau_n(u)\uparrow u$, $0\leq\tau_n(u)\leq u$ and $\tau_n(u)\in\{0,1/2^n,2/2^n,...,n\}$. Choose any $c\in\mathbb D$. Define $X_n=c+\text{sgn}(X-c)\tau_n\left(|X-c|\right)$. Then $X_n$ is simple. If $X<c$, then $X_n=c-\tau_n\left(|X-c|\right)$, so $X_n\downarrow X$. And if $X\geq c$, then $X_n=c+\tau_n\left(|X-c|\right)$, so $X_n\uparrow X$. And $X_n$ is between $X$ and $c$. Thus $\{X_n\}$ satisfies (RC). Hence $\rho(X_n)\to\rho(X)$. In addition, $X_n$ is simple, so $\rho(X_n)=\E[f(X_n)]-f(\E[X_n])$ and, by an argument similar to that in $(ii)\Rightarrow(i)$, we have $\E[f(X_n)]-f(\E[X_n])\to\E[f(X)]-f(\E[X])$. Therefore, we obtain $\rho(X)=\E[f(X)]-f(\E[X])$ for every $X\in\mathcal X_1({\mathbb D})$.
\end{proof}
\section{Proofs of  results in Section~\ref{s:identi_f}}\label{app:proof_identi_f}

% \subection{Proof of Lemma~\ref{le:equi_si_hda}}\label{app:proof_equi_si_hda}
\begin{proof}[Proof of Lemma~\ref{le:equi_si_hda}]
   The implication of (ii)$\Rightarrow$(i) is immediate. 
   
   (i)$\Rightarrow$(ii): If $\rho(X)=0$ for all such $X$, then $(ii)$ holds trivially. Suppose that $\rho$ is not identically $0$. For $m>0$, let $\mathcal P_m=\{\mu_X: X\in\mathcal X_s(\R_+)\text{ with } \E[X]=m\}$. For any $\mu\in\mathcal P_m$, define $U_m(\mu)=\int f(x)\mu(d x)-f(m)$. Note that $\mathcal P_m$ is a convex set, and $U_m(\mathcal P_m)$ is also a convex set on $\R$. Thus $U_m(\mathcal P_m)$ is an interval. We first show that $U_m(\mathcal P_m)\neq\{0\}$.
    % Now we want to show that $U_m(\mathcal P_m)$ is not identically $0$ on $\mathcal P_m$. 
    Since $\rho$ is not identically $0$, there exists $X_0\in\mathcal X_s(\R_+)$ such that $\rho(X_0)\neq 0$. Let $m_0=\E[X_0]$. Choose $\lambda_0\in (0,1)$ such that $n=(m-\lambda_0 m_0)/(1-\lambda_0)>0$. Let \(X_1\) and \(X_2\) be simple random variables with laws $\mu_{X_1}=\lambda_0\delta_{m_0}+(1-\lambda_0)\delta_n$ and $\mu_{X_2}=\lambda_0 \mu_{X_0}+(1-\lambda_0)\delta_n$ respectively. Then $\E[X_1]=\E[X_2]=m$ and $U_m(\mu_{X_2})-U_m(\mu_{X_1})=\lambda_0 U_{m_0}(\mu_{X_0})\neq 0$. Hence at least one of $U_m(\mu_{X_1})$ and $U_m(\mu_{X_2})$ is not $0$. Note that $U_m(\delta_m)=0$. Therefore, we have that $U_m(\mathcal P_m)$ is an interval with positive length.  Let $V_{c,m}(\mu)=\int f(cx)\mu(\d x)-f(cm)$. By (SI), we obtain that $U_m(\mu)\geq U_m(\nu) \Leftrightarrow V_{c,m}(\mu)\geq V_{c,m}(\nu)$ for $\mu,\nu\in\mathcal P_m$ and $c>0$. Therefore, there exists an increasing function $h_{c,m}$ such that $V_{c,m}(\mu)=h_{c,m}(U_m(\mu))$. For any $\mu,\nu\in \mathcal P_m$ and $\lambda\in[0,1]$, we obtain
   \begin{equation*}
   \begin{aligned}
       h_{c,m}(\lambda U_m(\mu)+(1-\lambda)U_m(\nu))&=V_{c,m}(\lambda\mu+(1-\lambda)\nu)\\&=\lambda h_{c,m}(U_m(\mu))+(1-\lambda)h_{c,m}(U_m(\nu)).
   \end{aligned}
   \end{equation*}
   Thus, $h_{c,m}(r)=a(c,m)r+b(c,m)$ for all $r\in U_m(\mathcal P_m)$ and some parameter functions $a,b$. Since $0=V_{c,m}(\delta_m)=h_{c,m}(U_m(\delta_m))=h_{c,m}(0)$, $b(c,m)=0$. Note that $h_{c,m}$ is increasing, so $a(c,m)\geq 0$. In addition, $V_{c,m}(\mathcal P_m)=U_{cm}(\mathcal P_{cm})$ for $c>0$, so we have that $V_{c,m}(\mathcal P_m)\neq\{0\}$. Hence, $a(c,m)>0$.

   Now we want to show that $a(c,m)$ does not depend on the value of $m$. Choose any $m_1,m_2>0$. We know that $U_m(\mathcal P_m)$ is an interval with positive length for any $m>0$, thus, there exists $\tilde X\in\mathcal X_s(\R_+)$ such that $\rho(\tilde X)\neq0$ and $\E[\tilde X]=m_1$. Choose $\tilde \lambda\in(0,1)$ and $\tilde n>0$ such that $m_2=\tilde\lambda m_1+(1-\tilde\lambda)\tilde n$. Define $\mu_{X_3}=\tilde\lambda\mu_{\tilde X}+(1-\tilde\lambda)\delta_{\tilde n}$ and $\mu_{X_4}=\tilde\lambda\delta  _{m_1}+(1-\tilde\lambda)\delta_{\tilde n}$. Then we have $U_{m_2}(\mu_{X_3})-U_{m_2}(\mu_{X_4})=\tilde\lambda U_{m_1}(\mu_{\tilde X})$. Similarly, we obtain $V_{c,m_2}(\mu_{X_3})-V_{c,m_2}(\mu_{X_4})=\tilde\lambda V_{c,m_1}(\mu_{\tilde X})$. Hence, we have $a(c,m_2) U_{m_1}(\mu_{\tilde X})=a(c,m_1)U_{m_1}(\mu_{\tilde X})$. Since $U_{m_1}(\mu_{\tilde X})=\rho(\tilde X)\neq0$, we have $a(c,m_1)=a(c,m_2)$ for any $m_1,m_2>0$. Thus $a(c,m)=A(c)>0$.

   Hence, we have $\rho(cX)=A(c)\rho(X)$ for all $X\in\mathcal X_s(\R_+)$. Note that $A(cd)\rho(X)=\rho(cdX)=A(c)\rho(dX)=A(c)A(d)\rho(X)$. Choose $X=X_0$ such that $\rho(X_0)\neq 0$, we have that $A(cd)=A(c)A(d)$ for all $c,d>0$. Since $f$ is measurable, $A(c)=\rho(cX_0)/\rho(X_0)$ is also measurable. Together with $A(c)>0$, we have that $A(c)=c^\alpha$ for some $\alpha\in\R$.
\end{proof}
% \subsection{Proof of Lemma~\ref{le:HDA_generator}}\label{app:rho_HDA_f}
\begin{proof}[Proof of Lemma~\ref{le:HDA_generator}]
% From Lemma~\ref{le:equi_si_hda}, we know that (SI) is equivalent to (HDA). Hence, we only need to prove the equivalence between (HDA) and the forms of $f$.\\
(ii)$\Rightarrow$(i): We can easily verify that (HDA) holds under $f_{a,\alpha}$.

(i)$\Rightarrow$(ii): The proof is similar to that of \citet[Theorem~3.11]{campbell2025excessgrowth}, in which they only show the case for $\alpha=0$. In the following, we derive $f$ for general $\alpha\in\R$.
% From (HDA), we obtain
% $
%     \mathbb E[f(cX)]-f(c\mathbb E[X)]=c^\alpha \mathbb E[f(X)]-c^\alpha f(\mathbb E[X)]
% $ holds for every $X\in\mathcal X_t^+$.
% Re-arranging the terms, we can have 
% $
%     \mathbb E[f(cX)]-c^\alpha \mathbb E[f(X)]=f\left(c\mathbb E[X]\right)-c^\alpha f(\mathbb E[X)].
% $
% Let \begin{equation}\label{eq:def_Psi_c}
%     \Psi_c(x)=f(cx)-c^\alpha f(x).
% \end{equation}
% From above equation, we have
% $
%    \mathbb E\left[\Psi_c(X)\right]=\Psi_c\left(\mathbb E[X]\right)
% $ for every $X\in\mathcal X_t^+$.
% % Let $X$ being a two-valued simple random variable taking ${x_1,x_2}$ with $\mathbb P(X=x_1)=t\in(0,1)$, then we have
% Thus we obtain
% $$
%     \Psi_c\left(tx_1+(1-t)x_2\right)=t\Psi_c(x_1)+(1-t)\Psi_c(x_2),
% $$
% where $x_1,x_2\in(0,\infty)$. Therefore, $\Psi_c$ is $t$-convex. Since $\Psi_c$ is measurable from the the measurability of $f$, we can have $\Psi_c$ is convex \citep[Corollary~3]{adamek2019remarks}. 
% % We obtain that $\Psi_c$ is $1/2$-convex \citep[Corollary~3]{nikodem2004tconvex}. 
% % Since $\Psi_c$ is measurable from the the measurability of $f$, we can have $\Psi_c$ is convex \citep{kotrys2013strongly}. 
% Similarly, we can obtain that $-\Psi_c$ is convex. 
From (HDA), we obtain that
$
    \E[f(cX)]-f(c\E[X])=c^\alpha \E[f(X)]-c^\alpha f(\E[X)]
$ holds for every $X\in\mathcal X_s(\R_+)$.
Rearranging the terms, we can have 
$
    \E[f(cX)]-c^\alpha \E[f(X)]=f\left(c\E[X]\right)-c^\alpha f(\E[X)].
$
Let \begin{equation}\label{eq:def_Psi_c}
    \Psi_c(x)=f(cx)-c^\alpha f(x).
\end{equation}
From the above equation, we have
$
   \E\left[\Psi_c(X)\right]=\Psi_c\left(\E[X]\right)$ for any $X\in\mathcal X_s(\R_+)$.
% Let $X$ being a two-valued simple random variable taking ${x_1,x_2}$ with $\mathbb P(X=x_1)=t\in(0,1)$, then we have
% Thus we obtain
% $$
%     \Psi_c\left(tx_1+(1-t)x_2\right)=t\Psi_c(x_1)+(1-t)\Psi_c(x_2),
% $$
% where $x_1,x_2\in(0,\infty)$. Therefore, $\Psi_c$ is $t$-convex. Since $\Psi_c$ is measurable from the the measurability of $f$, we can have $\Psi_c$ is convex \citep[Corollary~3]{adamek2019remarks}. 
% We obtain that $\Psi_c$ is $1/2$-convex \citep[Corollary~3]{nikodem2004tconvex}. 
% Since $\Psi_c$ is measurable from the the measurability of $f$, we can have $\Psi_c$ is convex \citep{kotrys2013strongly}. 
% Similarly, we can obtain that $-\Psi_c$ is convex. 
Hence, $\Psi_c$ is affine, that is
$
    \Psi_c(x)=\phi_1(c)x+\phi_2(c),
$
for some functions $\phi_1$ and $\phi_2$. Together with the definition of $\Psi_c$ in Eq.~\eqref{eq:def_Psi_c}, we obtain
\begin{equation}\label{eq:Psi_c_f}
    f(cx)=c^\alpha f(x)+\phi_1(c)x+\phi_2(c).
\end{equation}
Applying Eq.~\eqref{eq:Psi_c_f} with \(c\) replaced by \(cd\), we obtain
$
    f(cdx)=(cd)^\alpha f(x)+\phi_1(cd)x+\phi_2(cd).
$
Also, we can have
\begin{equation}\label{eq:f_2}
\begin{aligned}
    f(cdx)&=c^\alpha f(dx)+\phi_1(c)dx+\phi_2(c)
    \\&=(cd)^\alpha f(x)+c^\alpha \phi_1(d)x+c^\alpha\phi_2(d)+d\phi_1(c)x+\phi_2(c),
\end{aligned}
\end{equation}
where the first equality follows by applying Eq.~\eqref{eq:Psi_c_f} with \(x\) replaced by \(dx\), while the second equality follows by applying Eq.~\eqref{eq:Psi_c_f} with \(c=d\). By symmetry, we obtain
$
f(cdx)
=(cd)^\alpha f(x)
+d^\alpha\phi_1(c)x
+d^\alpha\phi_2(c)
+c\phi_1(d)x
+\phi_2(d).
$
Therefore, we obtain
\begin{equation}\label{eq:phi_1_2}
    \begin{cases}
    \phi_1(cd)=c^\alpha \phi_1(d)+\phi_1(c)d=d^\alpha\phi_1(c)+c\phi_1(d),\\
    \phi_2(cd)=c^\alpha\phi_2(d)+\phi_2(c)=d^\alpha\phi_2(c)+\phi_2(d).
\end{cases}
\end{equation}
For the case that $\alpha\neq 0$ and $\alpha\neq1$, choose $d\in(0,\infty)$ such that $d-d^\alpha\neq0$ and $1-d^\alpha\neq0$. 
% This is doable since from the definition of $\mathcal D_\alpha$, we know $$\mathcal D_\alpha\in\{(0,\infty),\,[0,\infty),\,(-\infty,0),\,(-\infty,0],\,\mathbb R\},$$ so $\mathcal C_\alpha$ is either the same as $\mathcal D_\alpha$ or $-\mathcal D_\alpha$. 
Then we can have $\phi_1(c)=\phi_1(d)(c-c^\alpha)/(d-d^\alpha)$, which indicates that $
    \phi_1(c)=k_1(c-c^\alpha)$
for some $k_1$. Similarly, $
    \phi_2(c)=k_2(1-c^\alpha)$
for some constant $k_2$. Hence, we obtain
$
   f(cx)=c^\alpha f(x)+ k_1(c-c^\alpha)x+k_2(1-c^\alpha).
$
% Hence, if $\mathcal D_\alpha\in\{(0,\infty),\,[0,\infty),\,\mathbb R\}$, then $\mathcal C_\alpha=\mathcal D_\alpha$. 
% If $\mathbb D_\alpha\in\{[0,\infty),\,(0,\infty),\,\R\}$, 
Taking $x=1$, we obtain that
$
    f(c)=\left(f(1)-k_1-k_2\right)c^\alpha+k_1 c+k_2,
$
for all $c\in(0,\infty)$. 
% If $\mathbb D_\alpha\in\{(-\infty,0],\,(-\infty,0)\}$, let $x=-1$, we can obtain $f(-c)=\left(f(-1)+k_1-k_2\right)(-c)^\alpha/(-1)^\alpha+k_1 (-c)+k_2$ for all $-c\in\mathbb D_\alpha$.
% Now if $D_\alpha\in\{(-\infty,0),\,(-\infty,0]\}$, then $C_\alpha=-\mathcal D_\alpha$. Hence we choose $x=-1$, and we obtain
% \begin{equation}\label{eq:form_f}
%     f(-c)=\frac{\left(f(-1)+k_1-k_2\right)}{(-1)^\alpha}(-c)^\alpha+k_1 (-c)+k_2,
% \end{equation}
% for all $-c\in\mathcal D_\alpha$. 
Therefore, for $\alpha\neq0,1$, we obtain
$f(x)=ax^\alpha+kx+b.$ 
% together with the convexity of $f$ on $\mathcal D_\alpha$, we have $a\beta\geq0$, as by design, $\beta x^\alpha$ is convex on $D_\alpha$.

For the case $\alpha=1$, Eq.~\eqref{eq:phi_1_2} becomes
\begin{equation}\label{eq:phi_1_2_alpha_1}
    \begin{cases}
    \phi_1(cd)=c \phi_1(d)+\phi_1(c)d=d\phi_1(c)+c\phi_1(d),\\
    \phi_2(cd)=c\phi_2(d)+\phi_2(c)=d\phi_2(c)+\phi_2(d).
\end{cases}
\end{equation}
Following the same procedures as $\alpha\neq0$ and $\alpha\neq1$ above, we can get
$
    \phi_2(c)=k_2(1-c)
$
for some constant $k_2$. As for $\phi_1$, we have $\phi_1(cd)=c \phi_1(d)+\phi_1(c)d$. Let $H(x)=\phi_1(x)/x$, then we obtain
\begin{equation}\label{eq:H}
    H(cd)=\frac{\phi_1(cd)}{cd}=\frac{c \phi_1(d)+\phi_1(c)d}{cd}=\frac{\phi_1(c)}c+\frac{\phi_1(d)}d=H(c)+H(d),
\end{equation}
which is the logarithmic functional equation. As $f$ is measurable on $(0,\infty)$, so is $H$. The measurable solution of the logarithmic functional equation in Eq.~\eqref{eq:H} is 
$
H(c)=a\log c.
$
Hence, $
    \phi_1(c)=ac\log c.
$
Putting the solutions of $\phi_1$ and $\phi_2$ back to Eq.~\eqref{eq:Psi_c_f}, we have 
$
    f(cx)=cf(x)+axc\log c+k_2(1-c).
$
Taking $x=1$, we obtain that the solution of the function $f$ under $\alpha=1$ is
\begin{equation}\label{eq:sol_f_alpha_1}
    f(c)=a c\log c+kc+b
\end{equation}
for $c\in(0,\infty)$, where $k,b\in \mathbb R$.

For the case that $\alpha=0$, Eq.~\eqref{eq:phi_1_2} becomes
\begin{equation}\label{eq:phi_1_2_alpha_0}
    \begin{cases}
    \phi_1(cd)= \phi_1(d)+\phi_1(c)d=\phi_1(c)+c\phi_1(d),\\
    \phi_2(cd)=\phi_2(d)+\phi_2(c)=\phi_2(c)+\phi_2(d).
\end{cases}
\end{equation}
For $\phi_1$, choosing $d=2$, we have 
\begin{equation}\label{eq:sol_phi_1_alpha_0}
    \phi_1(c)=k_1(1-c),
\end{equation}
for some constant $k_1$. Similarly, $\phi_2$ is the measurable solution of the logarithmic functional equation, thus $
    \phi_2(c)=a\log c$.
Therefore, taking $x=1$ in Eq.~\eqref{eq:Psi_c_f}, we have that the solution of $f$ under $\alpha=0$ is
\begin{equation}\label{eq:sol_f_alpha_0}
    f(c)=a\log c+kc+b,
\end{equation}
for $c\in(0,\infty)$. Note the affine part $kx+b$ is eliminated in the Jensen gap, that is, $\E[kX+b]-(k\E[X]+b)=0$, so we only need the nonlinear part of $f$ in the Jensen gap.
\end{proof}
% \subsection{Proof of Theorem~\ref{thm:variance_characterization}}\label{app:variance_chara_proof}
\begin{proof}[Proof of Theorem~\ref{thm:variance_characterization}]
   The implications (vi)$\Rightarrow$(i)--(v) are immediate. By Theorem~\ref{thm:rho_form_nonne_convex}, each of (i)--(v) gives $\rho=J_f$
for some convex $f:\R\to\R$. 
(i)$\Rightarrow$(vi) is from Lemma~\ref{le:ET_form_f}, and (ii)$\Rightarrow$(vi) is from Remark~\ref{re:large_domain_f2}.

(iii)$\Rightarrow$(vi):  
Since (ATI) implies (SI) and (TI),
Remark~\ref{re:large_domain_f} and
Lemma~\ref{le:TI_ET} give
$\rho(-Y)=\rho(Y)$ and
$\rho(Y+b)=e^{\beta b}\rho(Y)$.
If $\rho= 0 $, the conclusion is immediate.
Otherwise, choose $X$ with $\rho(X)>0$. Then
\[
e^\beta\rho(X)
=\rho(-X+1)
=\rho(X-1)
=e^{-\beta}\rho(X).
\]
Hence $\beta=0$, so (LocI) holds and
(i)$\Rightarrow$(vi) completes the argument.

%     (iii)$\Rightarrow$(vi): 
%     % Applying (LI), (BD), (N) with condition (HDA$_2$) or (LocI) is trivial to obtain $(ii)$ through Lemma~\ref{le:HDA_generator} or \ref{le:ET_form_f}. 
%     By Theorem~\ref{thm:rho_form_nonne_convex}, we know that $\rho$ is a Jensen gap with some convex generator for real-valued simple random variables. By (ATI), we know that $\rho$ satisfies both scaling invariance (SI) and translation invariance (TI) for $c\in\R$. By Remark~\ref{re:large_domain_f}, we obtain that $\rho(cX)=|c|^\alpha\rho(X)$ for every $c\neq0$ and $X\in\mathcal X_s(\R)$. Besides, by Theorem~\ref{thm:TI_form_f}, we know that the generator $f$ should take the form in \eqref{eq:TI_f} up to some affine term. Suppose that the exponential case holds and $\rho$ is not constantly $0$, so
% $f(x)=ae^{\beta x}+k'x+b'$ with $\beta\neq0$ and $a>0$. Together with $\rho(cX)=|c|^\alpha\rho(X)$, we can obtain that $\E[\psi(X)]=\psi(\E[X])$ where $\psi(x)=e^{c\beta x}-|c|^\alpha e^{\beta x}$. Hence, $\psi$ is affine in $x$ for every $c\neq0$. This is impossible when $\beta\neq0$. Hence, the exponential
% case is ruled out, and $f(x)=ax^2+k'x+b'$ for some $a\geq0$ and $k',b'\in\R$. Thus $\rho$ is the variance up to some nonnegative multiplicative constant.

(v)$\Rightarrow$(iv) is immediate.
(iv)$\Rightarrow$(vi): As shown in the proof of Proposition~\ref{prop:JG-convex}, quasi-convexity implies (LocI), which further implies that \(\rho=a\var\) for some \(a\geq0\).
% Fix $X_0\in\mathcal X_s(\R)$. Define $g(c)=\rho(X_0+c)$. It can be easily verified that $g$ is convex on $\R$. By (Cx) and note that $\rho(m)=0$ for $m\in R$, $g(c)=\rho\left((2X_0+2c)/2\right)\leq \rho(2X_0)/2$. Thus, $g$ is a bounded convex function on $\R$, which means that $g$ is a constant. Hence, $g(c)=\rho(X_0+c)=g(0)=\rho(X_0)$ for $c\in\R$ and  $X_0\in\mathcal X_s(\R)$, which is (LocI). Therefore, $\rho=a\var$ for $a\geq0$.
\end{proof}
\section{Proofs of results in Section~\ref{s:cond_fun_general_G}}\label{app:proof_gener_G}

We first derive  the formula  \eqref{eq:con_ES_finite}  in Example~\ref{ex:con_ES_G_finite} of Section \ref{s:con_finite}, which is a discrete version of Example~\ref{prop:con_ES}. 
\begin{proof}[Derivation of \eqref{eq:con_ES_finite}]
Let $\nu(X)=\operatorname{ES}_\alpha(X)=\min_{m\in\mathbb R}
\left\{
m+\int (x-m)^+\mu_X(\d x)/(1-\alpha)\right\}$. Therefore, from our Definition~\ref{def:con_risk_meas_fini}, its conditional version is 
\begin{equation*}
    \begin{aligned}
        \nu(X\mid \mG)=\ES_\alpha(X\mid \mG)&=\sum_{i\in I_+} \id_{A_i}\min_{m\in\mathbb R}
\left\{
m+\frac{1}{1-\alpha}\int (x-m)^+\mu_{X\mid A_i}(\d x)\right\}
% \\&=\sum_{i=1}^n 
% \left(
% m^*_i+\frac{1}{1-\alpha}\int (x-m^*_i)^+\mu_{X\mid A_i}(\d x)\right)\id_{A_i}
\\&=\sum_{i\in I_+}
\left(
m^*_i+\frac{1}{1-\alpha} \E[(X-m^*_i)^+\mid A_i]\right)\id_{A_i}\text{ a.s.,}
    \end{aligned}
\end{equation*}
 % where $m^*_i=\inf\{t\in\mathbb R:\mu_{X\mid A_i}((-\infty,t])\ge\alpha\}$ \citep[Definition~1 and Theorem~10]{rockafellar_uryasev_2002}. 
 where $m^*_i$ is a minimizer of $\min_{m\in\mathbb R}
\left\{
m+\int (x-m)^+\mu_{X\mid A_i}(\d x)/(1-\alpha)\right\}$. For any $Z\in L^0(\mG)$, here $L^0(\mG)$ refers to the set of all $\mG$-measurable random variables, we have $Z=\sum_{i\in I_+} \id_{A_i} z_i$ a.s., thus we obtain
\begin{equation*}
\begin{aligned}
        Z+\frac1{1-\alpha}\E[(X-Z)^+\mid \mG]&=\sum_{i\in I_+}
\left(
z_i+\frac{1}{1-\alpha} \E[(X-z_i)^+\mid A_i]\right)\id_{A_i}\\&\geq\sum_{i\in I_+}
\left(
m^*_i+\frac{1}{1-\alpha} \E[(X-m^*_i)^+\mid A_i]\right)\id_{A_i}=\ES_\alpha(X\mid \mG)\quad\text{a.s.},
\end{aligned}
\end{equation*}
where the first equality follows from Example~\ref{ex:con_E_X_Y_finite}. Therefore
\begin{equation*}
    \ES_\alpha(X\mid\mG)\leq \essinf_{Z\in L^0(\mG)}\left\{Z+\frac1{1-\alpha}\E[(X-Z)^+\mid \mG]\right\}\quad\text{a.s.}
\end{equation*}
Besides, let $Y=\sum_{i\in I_+} \id_{A_i} m^*_i$. Obviously, $Y\in L^0(\mG)$. Thus, we obtain
\begin{equation*}
\begin{aligned}
    \ES_\alpha(X\mid\mG)&=Y+\frac1{1-\alpha}\E[(X-Y)^+\mid \mG]\\&\geq\essinf_{Z\in L^0(\mG)}\left\{Z+\frac1{1-\alpha}\E[(X-Z)^+\mid \mG]\right\}\quad\text{a.s.}
\end{aligned}
\end{equation*}
The two inequalities yield \eqref{eq:con_ES_finite}.
\end{proof} 
% \subsection{Proof of Proposition~\ref{thm1:measurable_con_func}}\label{app:measure_con_func}
\begin{proof}[Proof of Proposition~\ref{thm1:measurable_con_func}]
Since \(\mathcal P\in\Sigma_{\mathcal M}\), we extend \(\tilde{\nu}\) measurably to \(\mathcal M(S)\) by setting it to \(0\) outside \(\mathcal P\).
Then \(\nu(X\mid\mathcal G)=\tilde{\nu}\circ\mu_{X\mid\mathcal G}\) a.s.  Therefore, it remains to prove that $\mu_{X|\mathcal G}:(\Omega,\mathcal G)\to(\mathcal M(S),\Sigma_{\mathcal M})$ is measurable. For any generator $e_B^{-1}(\tilde{B})$ of $\Sigma_{\mathcal M}$, we have
\begin{equation*}
\begin{aligned}
    \mu_{X|\mathcal G}^{-1}(e_B^{-1}(\tilde{B}))&=\{\omega:\mu_{X|\mathcal G}(\omega,\cdot)\in e_B^{-1}(\tilde{B})\}=\{\omega:\mu_{X|\mathcal G}(\omega,B)\in \tilde{B}\}
\end{aligned}
\end{equation*}
Note that $\mu_{X\mid\mathcal G}(\cdot,B)$ is measurable with respect to $\mathcal G$, hence $\mu_{X|\mathcal G}^{-1}(e_B^{-1}(\tilde{B}))\in \mathcal G$. Besides, we can easily check that the set $\Sigma'=\{S':\mu_{X|\mathcal G}^{-1}(S')\in\mathcal G\}$ is a $\sigma$-algebra, then $\Sigma_{\mathcal M}\subseteq\Sigma'$. Hence, for any element $\tilde{S}\in\Sigma_{\mathcal M}$, we have $\mu_{X|\mathcal G}^{-1}(\tilde{S})\in \mathcal G$, which means that $\mu_{X|\mathcal G}$ is measurable.
\end{proof}
% \subsection{Proof of Lemma~\ref{le:mean_meas}}\label{app:mean_meas}
\begin{proof}[Proof of Lemma~\ref{le:mean_meas}]
The measurability follows the standard extension argument for the Lebesgue integral: we first establish this for indicator functions, then for nonnegative simple functions, then for nonnegative Borel functions by the monotone convergence theorem, and finally for general integrable functions by decomposing $f=f^+-f^-$, where $f^+=\max\{f,0\}$ and $f^-=\max\{-f,0\}$.
For the nonnegative $f$, we can also prove that $\tilde{\nu}_f(\mu)$ is measurable with respect to $\Sigma_{\mathcal M}$ under the extended sense, that is, we allow $\tilde{\nu}_f(\mu)$ to take value $\infty$. Thus, if we restrict the domain to be $D_f$, it is also measurable.
% First, let $f=1_B$ for some $B\in\mathcal B(S)$. Then
% $\tilde{\nu}_f(\mu)=\int 1_B\,\text{d}\mu=\mu(B)=e_B(\mu)$,
% which is $\Sigma_{\mathcal M}$-measurable by the definition of $\Sigma_{\mathcal M}$.
% Next, let $f=\sum_{k=1}^n a_k1_{B_k}$ be a nonnegative simple function, where $a_k\ge0$ and $B_k\in\mathcal B(S)$. Then
% $\int f\,\text{d}\mu=\sum_{k=1}^n a_k\mu(B_k)=\sum_{k=1}^n a_k e_{B_k}(\mu)$.
% Hence $\mu\mapsto\int f\,d\mu$ is $\Sigma_{\mathcal M}$-measurable.
% Now let $f:S\to[0,\infty]$ be nonnegative and Borel measurable. Choose nonnegative simple functions $f_n$ such that $f_n\uparrow f$. By the monotone convergence theorem,
% $\int f_n\,\text{d}\mu\uparrow\int f\,\text{d}\mu$
% for every $\mu\in\mathcal M(S)$. Since each map $\mu\mapsto\int f_n\,\text{d}\mu$ is measurable, the pointwise limit
% $\mu\mapsto\int f\,\text{d}\mu$
% is $\Sigma_{\mathcal M}$-measurable as an $[0,\infty]$-valued function. Therefore, if we restrict the domain to be $D_f$, it is also measurable.
Once measurability has been established for nonnegative $f$, we apply the result to $|f|$. Therefore, $D_f=\{\mu\in\mathcal M(S):\int |f|\,\d\mu<\infty\}=\bigcup_{m=1}^{\infty}\{\mu\in\mathcal M(S):\int |f|\,\d\mu\leq m\}\in \Sigma_{\mathcal M}.$
% \begin{align*}
%     D_f&=\{\mu\in\mathcal M(S):\int |f|\,\d\mu<\infty\}=\bigcup_{m=1}^{\infty}\{\mu\in\mathcal M(S):\int |f|\,\d\mu\leq m\}\in \Sigma_{\mathcal M}.
% \end{align*}
% Finally, for a general real-valued Borel function $f$, write $f=f^+-f^-$. On $D_f$, $\int f\,\text{d}\mu$ is well-defined. By the nonnegative case, the maps
% $\mu\mapsto\int f^+\,\text{d}\mu$ and $\mu\mapsto\int f^-\,\text{d}\mu$
% are measurable. Hence their difference
% $\mu\mapsto\int f\,\text{d}\mu$
% is measurable from $(D_f,\Sigma_{\mathcal M}|_{D_f})$ to $(\mathbb R,\mathcal B(\mathbb R))$.
% If $\mu_X$ is the law of $X$, it is trivial that
% $\tilde{\Gamma}_f(\mu_X)=\int f\,d\mu_X=\mathbb E[f(X)]$.
\end{proof}
% \subsection{Proof of Proposition~\ref{prop:con_mean}}\label{app:con_mean}
\begin{proof}[Proof of Proposition~\ref{prop:con_mean}]
The identity $\tilde{\nu}_f(\mu_{X\mid\mathcal G})
=\int f\,\d\mu_{X\mid\mathcal G}
=\E[f(X)\mid\mathcal G]\text{ a.s.}$ can be established first for indicator functions, then for nonnegative simple functions, then for nonnegative Borel functions, and finally for general integrable functions. It remains to verify that $\mu_{X\mid\mathcal G}(\omega,\cdot)\in D_f$ a.s.
If $\mu_X\in D_f$, then $\E\left[\int |f|\,\d\mu_{X\mid\mathcal G}\right]=\E\left[|f(X)|\right]=\int |f|\,\d\mu_X<\infty$. Since $\int |f|\,\d\mu_{X\mid\mathcal G}\geq 0$, it follows that $\int |f|\,\d\mu_{X\mid\mathcal G}<\infty$ a.s. Hence, $\mu_{X\mid\mathcal G}(\omega,\cdot)\in D_f$ a.s.
\end{proof}
\begin{proof}[Proof of the equalities in Example~\ref{prop:con_ES}]
 By \eqref{eq:es_def},
\[
\operatorname{ES}_\alpha(X)
    =
    \inf_{q\in\mathbb Q}
    \left\{
        q+\frac{1}{1-\alpha}\int (x-q)^+\,\mu_X(\d x)
    \right\},\]
where $\mu_X$ is the law of $X$. This means that $\operatorname{ES}$ is measurable as a function of the law. By Proposition~\ref{prop:con_mean}, $K_\omega$ has finite first moment
almost surely.
Hence, by Definition~\ref{def:con_fun},
\begin{equation}\label{eq:ES_min_equa}
    \operatorname{ES}_\alpha(X\mid\mathcal G)(\omega)
    =
    \min_{m\in\mathbb R}
    \left\{
        m+\frac{1}{1-\alpha}
        \int (x-m)^+\,K_\omega(\d x)
    \right\}.
\end{equation}
By \citet[Definition~1 and Theorem~10]{rockafellar_uryasev_2002}, a minimizer in \eqref{eq:ES_min_equa} is
$$  q_\alpha(\omega)
    =
    \inf\{t\in\mathbb R:K_\omega((-\infty,t])\geq\alpha\}.$$
Moreover, $q_\alpha$ is $\mathcal G$-measurable, since for every
$r\in\mathbb R$,
\[
    \{\omega:q_\alpha(\omega)<r\}
    =
    \bigcup_{\substack{a\in\mathbb Q\\a<r}}
    \{\omega:K_\omega((-\infty,a])\geq\alpha\}
    \in\mathcal G.
\]
On the set $\{q_\alpha>0\}$, $\E[X^+\mid\mathcal G](\omega)
=
\int x^+\,K_\omega(\d x)
\ge
q_\alpha(\omega) K_\omega([q_\alpha(\omega),\infty))
\ge
(1-\alpha)q_\alpha(\omega)
$ a.s.,
$
q_\alpha^+
\le
\frac{1}{1-\alpha}\E[X^+\mid\mathcal G]
$ a.s.
Similarly, we have $ q_\alpha^-
    \leq
    \frac{1}{\alpha}\E[X^-\mid\mathcal G]$ a.s.
%     on the set $\{q_\alpha<0\}$,
% $$
% \mathbb E[X^-\mid\mathcal G]
% =
% \int x^-\,K_\omega(\text{d}x)
% \ge
% (-q_\alpha)K_\omega((-\infty,q_\alpha])
% \ge
% \alpha(-q_\alpha)\quad \text{a.s.}
% $$
% Its integrability follows from
% \[
%     q_\alpha^+
%     \leq
%     \frac{1}{1-\alpha}\E[X^+\mid\mathcal G],
%     \qquad
%     q_\alpha^-
%     \leq
%     \frac{1}{\alpha}\E[X^-\mid\mathcal G],
% \]
Hence $q_\alpha\in L^1(\mathcal G)$.
By the disintegration formula in
\citet[Theorem~8.5]{kallenberg2021}, for every
$Z\in L^1(\mathcal G)$,
\(
    \int (x-Z(\omega))^+\,K_\omega(\d x)
    =
    \E[(X-Z)^+\mid\mathcal G](\omega)
\) a.s.
Therefore, the minimizing property of $q_\alpha$ yields
\[
    \operatorname{ES}_\alpha(X\mid\mathcal G)= q_\alpha+\frac{1}{1-\alpha}\E[(X-q_\alpha)^+\mid\mathcal G]
    \leq
    Z+\frac{1}{1-\alpha}\E[(X-Z)^+\mid\mathcal G]
    \quad\text{a.s.}
\]
for every $Z\in L^1(\mathcal G)$. Taking the essential infimum gives
one inequality, while the reverse inequality follows by choosing
$Z=q_\alpha$.
\end{proof}
\begin{proof}[Proof of Proposition~\ref{cor:dynamic_jensen}]
Write $M_t=\E[X\mid\mathcal F_t]$. Since $\mathbb D$ is an open interval, $M_t\in\mathbb D$ a.s.; at each finite endpoint, this follows from the strict positivity of the conditional expectation of the distance of $X$ from that endpoint. Choose an affine minorant $\ell$ of $f$ and put $g=f-\ell\geq0$. Then $g(X)\in L^1$ and conditional Jensen's inequality gives
\[
0\leq g(M_t)\leq\E[g(X)\mid\mathcal F_t].
\]
Thus $g(M_t)$ and $f(M_t)$ are integrable. Conditional Jensen's inequality also gives $D_t(X)\geq0$, and its definition shows that $D_t(X)\in L^1$. Finally, for $s\leq t$, the tower property yields
\[
\begin{aligned}
\E[D_t(X)\mid\mathcal F_s]+D_s(M_t)
&=\E[f(X)\mid\mathcal F_s]-f(M_s) =D_s(X).
\end{aligned}
\]
The supermartingale property follows from $D_s(M_t)\geq0$.
\end{proof} 
% \subsection{Proof of Lemma~\ref{le:law_domain_meas}}\label{app:law_domain_meas}
\begin{proof}[Proof of Lemma~\ref{le:law_domain_meas}]
% $\mathcal X_s(\mathbb D)$ in Theorems~\ref{thm:rho_form_simple}, \ref{thm:rho_form_nonne_convex}, \ref{thm:log_characterization} and \ref{thm:variance_characterization},
% bounded random variables $\mathcal X_c(\tilde{\mathbb D})$ in Proposition~\ref{thm:conti}, and integrable random variables $\mathcal X_1(\tilde{\mathbb D})$ in Proposition~\ref{thm:L1_BD}.
For the law domain $\mathcal P_s$ corresponding to $\mathcal X_s(\mathbb D)$ with an interval $\mathbb D$ in Theorems~\ref{thm:rho_form_simple}, \ref{thm:rho_form_nonne_convex}, \ref{thm:log_characterization} and \ref{thm:variance_characterization}, let
\(
B_r=\mathbb D\cap(-\infty,r].
\)
Let $\Q$ denote the set of rational numbers. For every $\mu\in\mathcal P_s$, the quantity
\(
e_{B_q}(\mu)=\mu(B_q)
\)
can take only finitely many distinct values as $q$ ranges over $\Q$. For $m\geq 1$, let
\[
\mathcal P_{s,m}
=
\left\{
\mu\in\mathcal M(\mathbb D):
\#\left\{
e_{B_q}(\mu):q\in\Q
\right\}\leq m
\right\}=\left\{
\mu\in\mathcal M(\mathbb D):
\#\left\{
e_{B_r}(\mu):r\in\R
\right\}\leq m
\right\},
\] where the last equality follows since, for every $r\in\R$, one can choose
$q_n\in\Q$ such that $q_n\downarrow r$. Hence $e_{B_r}(\mu)=\lim_{q_n\downarrow r} e_{B_{q_n}}(\mu)$.
Note that, for every $\mu\in\mathcal P_{s,m}$ and every distinct
$q_1,\ldots,q_{m+1}\in\Q$, at least two of the values
$e_{B_{q_1}}(\mu),\ldots,e_{B_{q_{m+1}}}(\mu)$ must coincide. Therefore,
\[
\mathcal P_{s,m}
=
\bigcap_{\substack{
q_1<\cdots<q_{m+1}\\
q_1,\ldots,q_{m+1}\in\Q
}}
\;
\bigcup_{1\leq i<j\leq m+1}
\left\{
\mu\in\mathcal M(\mathbb D):
e_{B_{q_i}}(\mu)
=
e_{B_{q_j}}(\mu)
\right\}.
\]
Since $e_{B_{q}}$ is measurable with respect to the evaluation $\sigma$-algebra $\Sigma_{\mathcal M}$,
% For fixed $q_i$ and $q_j$, the set
% \[
% \left\{
% \mu\in\mathcal M(\mathbb D):
% e_{B_{q_i}}(\mu)=e_{B_{q_j}}(\mu)
% \right\}
% \]
% belongs to $\Sigma_{\mathcal M}$, since both $e_{B_{q_i}}$ and $e_{B_{q_j}}$ are measurable with respect to the evaluation $\sigma$-algebra. As only countable unions and intersections are involved, $\mathcal P_{s,m}\in\Sigma_{\mathcal M}$. Therefore,
\(
\mathcal P_s
=
\bigcup_{m=1}^{\infty}\mathcal P_{s,m}
\in\Sigma_{\mathcal M}.
\)

For the law domain $\mathcal P_c$ corresponding to $\mathcal X_c({\mathbb D})$ with an open interval $\mathbb D$ in Proposition~\ref{thm:conti}, we have
\[
\mathcal P_c
=
\bigcup_{\substack{
a,b\in\Q\\
[a,b]\subseteq{\mathbb D}
}}
\left\{
\mu\in\mathcal M({\mathbb D}):
\mu([a,b])=1
\right\}
=
\bigcup_{\substack{
a,b\in\Q\\
[a,b]\subseteq{\mathbb D}
}}
e_{[a,b]}^{-1}(\{1\})\in\Sigma_{\mathcal M}.
\]
% Since each $e_{[a,b]}$ is measurable and there are only countably many rational pairs $(a,b)$, it follows that
% \[
% \mathcal P_c\in\Sigma_{\mathcal M},
% \]

% For every fixed $q\in\Q$,
% \[
% \left\{
% \mu\in\mathcal M(\mathbb D):
% e_{B_q}(\mu)\in G_n
% \right\}
% =
% e_{B_q}^{-1}(G_n)
% \in\Sigma_{\mathcal M},
% \]
% since $e_{B_q}$ is measurable and $G_n$ is a finite Borel subset of $[0,1]$. Hence,
% \[
% \mathcal P_e\in\Sigma_{\mathcal M}.
% \]

Finally, for the law domain $\mathcal P_1$ corresponding to $\mathcal X_1({\mathbb D})$ with an open interval $\mathbb D$ in Proposition~\ref{thm:L1_BD}, its measurability has already been established in Lemma~\ref{le:mean_meas}. Indeed, applying Lemma~\ref{le:mean_meas} with $f(x)=x$ gives
\[
\mathcal P_1
=
\left\{
\mu\in\mathcal M({\mathbb D}):
\int_{{\mathbb D}} |x|\,\mu(dx)<\infty
\right\}
\in\Sigma_{\mathcal M}.
\]
Therefore, all the law domains appearing in the characterization theorems of Section~\ref{s:jensen_func} are measurable with respect to their corresponding evaluation $\sigma$-algebras.

Hence, we have verified that the law domains $\mathcal P$ appearing in the characterization results in Section~\ref{s:jensen_func} are measurable. This ensures that, together with the assumed measurability of the law representation $\tilde\rho$, the conditional functional $\rho(X\mid\mG)$ is measurable whenever the conditional law remains in the relevant law domain. This is because, by Proposition~\ref{thm1:measurable_con_func},
$\mu_{X\mid\mG}$ is $\mG$-measurable as an
$(\mathcal M,\Sigma_{\mathcal M})$-valued mapping. Hence, if
$\mathcal P\in\Sigma_{\mathcal M}$ and $\mu_{X\mid\mG}$ takes values in $\mathcal P$, then $\mu_{X\mid\mG}$ is also $\mG$-measurable
as a $(\mathcal P,\Sigma_{\mathcal M}|_{\mathcal P})$-valued mapping.
Therefore, since $\tilde\rho$ is measurable on
$(\mathcal P,\Sigma_{\mathcal M}|_{\mathcal P})$,
$\rho(X\mid\mG)=\tilde\rho(\mu_{X\mid\mG})$ is $\mG$-measurable.
\end{proof}

\section{Star-shaped Jensen gaps}
\label{app:JG-star}

Following \citet{CCMTW22}, a functional $\rho:L^q\to\RR$
with $\rho(0)=0$ is called \emph{star-shaped} if
$\rho(tX)\le t\rho(X)$ for all $X\in L^q$ and $t\in(0,1)$.

\begin{proposition}
Let $q\in[1,\infty]$, and let $f:\R\to\R$ be measurable
such that $J_f$ is well-defined on $L^q$.
Then $J_f$ is star-shaped if and only if
$g_t:x \mapsto tf(x)-f(tx)$ is convex on $\R$ for every $t\in(0,1)$.
\end{proposition}

\begin{proof}
For simple $X$, we have
$J_{g_t}(X)=tJ_f(X)-J_f(tX)$.
Thus necessity follows by testing two-point random variables.
Conversely, fix $t\in(0,1)$ and $X\in L^q$, put $m=\E[X]$,
and choose a supporting line
$g_t(x)\ge g_t(m)+a(x-m)$ for all $x\in \R$.
Rearranging gives
\[
f(tx)-f(tm)
\le t\bigl(f(x)-f(m)\bigr)-a(x-m).
\]
Taking expectations with $x$ replaced by $X$ yields $J_f(tX)\le tJ_f(X)$,
also for extended values.
\end{proof}

Star-shapedness is strictly weaker than convexity or
quasiconvexity of $J_f$, even among nonnegative Jensen gaps.
Indeed, $f(x)=|x|$ gives $J_f(tX)=tJ_f(X)$,
but $J_f$ is not a multiple of variance.
Convexity of $f$ does not imply star-shapedness of $J_f$:
for $f(x)=(x+1)^+$ and $X$ taking values $-4,0$ with equal
probabilities,
 $
J_f(X/2)=J_f(X)=1/2>J_f(X)/2.
 $ 
Finally, $f(x)=-|x|$ gives a star-shaped Jensen gap
which is not nonnegative.

\bibliography{rho_form}
 
%    \begin{thebibliography}{10}
 
% \end{thebibliography} 

\end{document}